\documentclass[11pt]{article}

\usepackage[preprint]{acl}

\usepackage{times}
\usepackage{latexsym}

\usepackage[T1]{fontenc}
\usepackage[utf8]{inputenc}

\usepackage{microtype}

\usepackage{inconsolata}

\usepackage{graphicx}

\usepackage{multirow}
\usepackage{array}

\usepackage{diagbox}
\usepackage{booktabs}
\usepackage{multirow}
\usepackage{graphicx}
\usepackage{amsmath}
\usepackage{array}
\usepackage{makecell}
\usepackage{subcaption}
\usepackage{enumitem}
\usepackage{hyperref}
\usepackage{tcolorbox}
\tcbuselibrary{skins,breakable,listings}
\usepackage{xcolor}
\usepackage{float}

\usepackage{tabularx}

\usepackage{seqsplit}

\usepackage{needspace}

\title{DataEvolver: Automatic Data Preparation for Large Language Models through Multi-Level Self-Evolving}
\renewcommand{\thefootnote}{\fnsymbol{footnote}}

\author{
  \textbf{Chao Deng}\footnotemark[1],\quad
  \textbf{Shaolei Zhang}\thanks{Equal contribution.},\quad
  \textbf{Ju Fan}\thanks{Corresponding author: Ju Fan, Xiaoyong Du},\quad
  \textbf{Xiaoyong Du}\footnotemark[2] \\
  Renmin University of China, Beijing, China \\
  \texttt{\{dengc, zhangshaolei98, fanj, duyong\}@ruc.edu.cn}
}

\newcommand{\hflink}[1]{\href{https://huggingface.co/datasets/#1}{\texttt{#1}}}

\newtcolorbox{seedbox}[2][]{breakable,enhanced,
  before={\par\needspace{12\baselineskip}},
  colback=black!2,colframe=black!25,boxrule=0.6pt,arc=1mm,
  left=1.2mm,right=1.2mm,top=1.0mm,bottom=1.0mm,
title=\textbf{#2},fonttitle=\small\bfseries,coltitle=black,
  attach title to upper,
  #1}

\newtcblisting{seedcode}[1][]{
  listing only,
  breakable,
  enhanced,
  colback=white,
  colframe=black!15,
  boxrule=0.4pt,
  arc=0.8mm,
  left=1mm,right=1mm,top=0.8mm,bottom=0.8mm,
  listing options={
    basicstyle=\ttfamily\scriptsize,
    columns=fullflexible,
    breaklines=true,
    breakatwhitespace=false,
    showstringspaces=false
  },
  #1
}

\newcommand{\PreserveBackslash}[1]{\let\temp=\\#1\let\\=\temp}
\newcolumntype{C}[1]{>{\PreserveBackslash\centering}p{#1}}
\newcolumntype{R}[1]{>{\PreserveBackslash\raggedleft}p{#1}}
\newcolumntype{L}[1]{>{\PreserveBackslash\raggedright}p{#1}}

\newcommand{\blue}[1]{\textcolor{black}{#1}}

\begin{document}
\maketitle
\setcounter{footnote}{0}
\renewcommand{\thefootnote}{\arabic{footnote}}

\begin{abstract}
High-quality training data is essential to large language models (LLMs) and typically requires extensive and costly manual curation.
\blue{Automatic data preparation systems support reusable operators and task-driven workflows, yet adapting these workflows to the desired data quality remains challenging. High-quality examples provide a practical specification for guiding this adaptation. We introduce DataEvolver, a data preparation system that automatically constructs pipelines through multi-level self-evolution. At the operator level, it expands the operator library and refines logical plans to resolve dependency conflicts. At the pipeline level, it instantiates these plans into code and uses trial-output feedback to refine the pipeline toward seed-specified quality requirements.}
\blue{Experiments on seven benchmarks show that DataEvolver substantially improves data quality and achieves an average gain of approximately 10 score points in downstream LLM performance compared with training on original data, highlighting new opportunities for the iterative co-evolution of LLMs and data.}
% The public repository link is retained here for restoration after anonymous review.
% \footnote{Code: \href{https://github.com/ruc-datalab/DataEvolver}{https://github.com/ruc-datalab/DataEvolver}.}
\end{abstract}

\section{Introduction}
\label{sec:intro}

Large language models (LLMs) have demonstrated strong performance across a wide range of tasks~\cite{openai2023gpt4technicalreport}, largely driven by the availability of high-quality training data~\cite{kaplan2020scalinglaws, hoffmann2022trainingcomputeoptimal, li2025datacomplmsearchgenerationtraining}. However, acquiring such data is very challenging, as raw data is often noisy and manual data preparation is prohibitively expensive~\cite{ouyang2022traininglanguagemodelstofollow,zhu2026surveydataagentsemerging}. As a result, automatic data preparation for LLMs has received increasing attention~\cite{zha2023datacentricai,chen2023datajuiceronestopdataprocessing,liang2025dataflowllmdrivenframeworkunified}.

The goal of data preparation is to orchestrate an effective pipeline that transforms raw data into high-quality target data~\cite{rahm2000data, krishnan2016alphaclean}. In practice, this process is typically carried out by human experts, who inspect the raw data, form a conceptual understanding of the desired target data, and design a corresponding transformation pipeline~\cite{ratner2017snorkel, kandel2011wrangler}. Ideally, an automatic system should emulate this process by flexibly constructing end-to-end pipelines based on the characteristics of the raw data, guided by a small set of high-quality examples or rules.

\begin{figure}[t]
	\centering
    \includegraphics[width=\columnwidth]{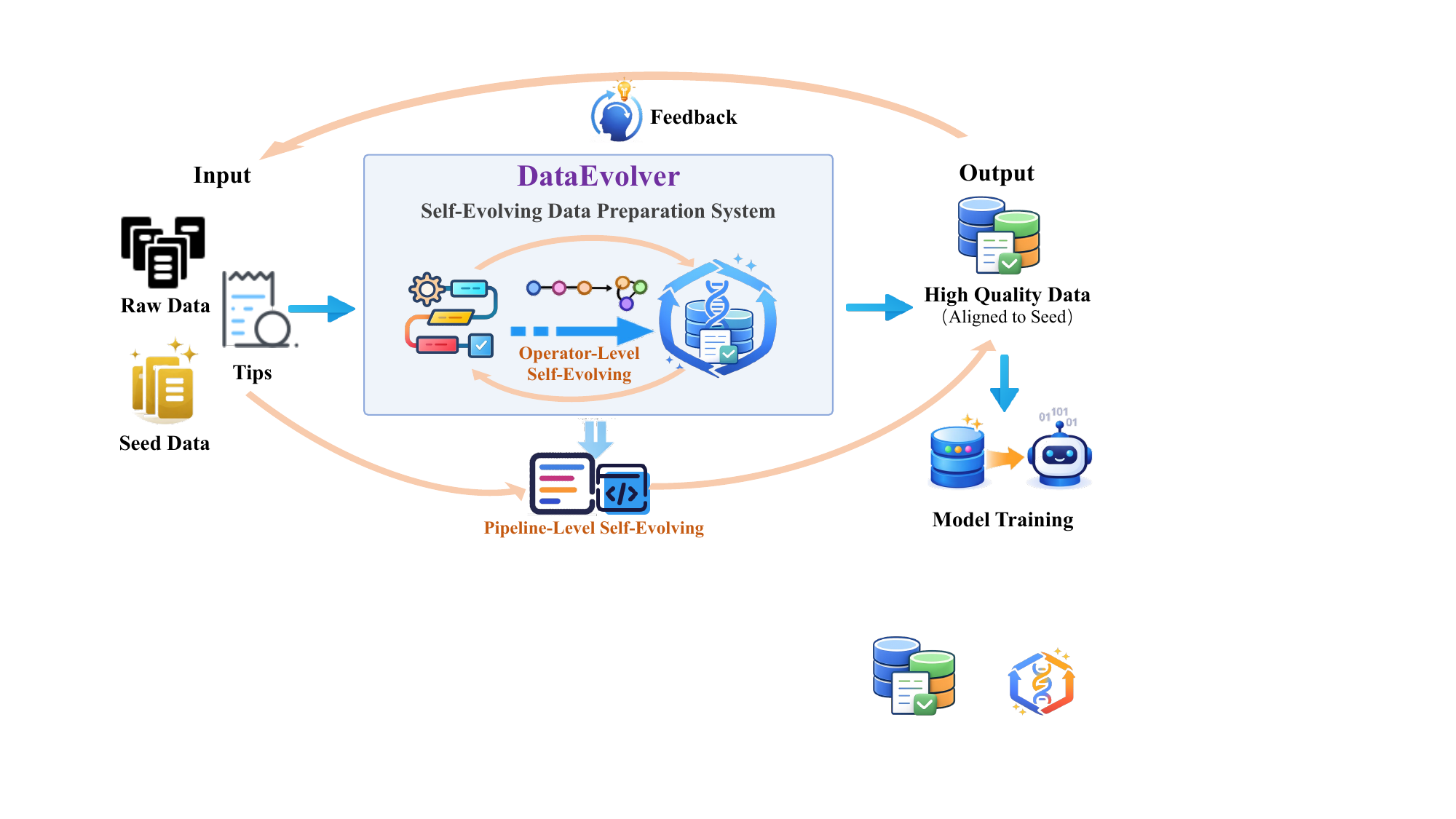}
    \caption{
    DataEvolver automatically transforms raw data into high-quality training data for LLMs through the proposed multi-level self-evolving.
    %Illustration of DataEvolver, which automatically transforms raw data into LLM-oriented high-quality data through multi-level self-evolving.
    }
	\label{fig:figure1}
    \vspace{-1em}
\end{figure}

\blue{Existing data preparation methods include operator-based systems and pipeline synthesis approaches. Operator-based systems provide reusable transformations and scalable execution~\cite{chen2023datajuiceronestopdataprocessing,chen2025datajuicer20cloudscaleadaptive}, while synthesis approaches construct and refine workflows from task specifications~\cite{liang2025dataflowllmdrivenframeworkunified,ge2025texttopipelinebridgingnaturallanguage,alidu2025prompt2dagmodularmethodologyllmbased,chang2025llapipellmguidedreinforcementlearning,10.14778/3750601.3750671,autoprep}. These advances make automatic orchestration increasingly practical. A key question is how to use high-quality examples to guide both the construction of executable pipelines and the refinement of their outputs.}

%In this paper, we aim to develop a fully automatic data preparation system. Given raw data and some high-quality target examples (denoted as seed data), the system is expected to construct a pipeline that transforms raw data into high-quality data. 

In this paper, we study the problem of flexible automatic data preparation under minimal supervision. Given raw data and a small set of high-quality target examples (referred to as \emph{seed data}), automatically constructing an appropriate transformation pipeline introduces two key challenges.
First, the transformation from raw data to target data often involves long chains of operations, making automatically generated pipelines prone to logical inconsistencies or execution failures (i.e., \emph{executability}). Second, even when a pipeline executes successfully, its outputs may still deviate from the desired quality, potentially degrading downstream model performance (i.e., \emph{effectiveness}).

To address these challenges, we propose DataEvolver, which couples two levels of self-evolution for automatic data preparation. Operator-level evolution adapts operators and their dependencies to improve logical executability, while pipeline-level evolution uses trial-output feedback to improve preparation effectiveness. Specifically, DataEvolver first infers a profile of high-quality data from raw data and seed data. Guided by this profile, it generates a directed acyclic graph (DAG) of logical operators as a transformation plan and iteratively evolves both the operator library and the plan. The logical plan is then instantiated into executable code and evaluated through trial runs on a sampled subset. Comparing trial outputs with the seed data produces experience that guides subsequent data understanding and pipeline construction. The refined pipeline is then applied to the full dataset.

Experiments on seven benchmarks show that DataEvolver substantially improves data quality, achieving average gains of approximately 10 score points over raw data and 2 score points over a strong data preparation system across multiple LLMs. Overall, DataEvolver recasts data preparation as the evolution of a reusable transformation pipeline guided by high-quality examples. Its multi-level design connects operator adaptation with pipeline refinement, providing a practical framework for automatic LLM data preparation and highlighting a path toward the iterative co-evolution of LLMs and data.

%To this end, we propose DataEvolver, the first self-evolving data preparation system that iteratively boosts data quality. DataEvolver employs a multi-level self-evolving framework to construct data preparation pipelines, where operator-level evolution ensures logical executability and pipeline-level evolution guarantees that the resulting data meets quality requirements. Specifically, DataEvolver understands the feature profile of high-quality data from raw and seed data. Guided by this understanding, DataEvolver generates a Directed Acyclic Graph (DAG) of logical operators as a plan and iteratively evolves the operator library and the plan. Then, the logical plan is instantiated into executable code and evaluated through trial runs on samples subset. By measuring discrepancies between trial outputs and the seed data, DataEvolver acquires new experience and accordingly evolves the entire pipeline. The evolution continues until data quality converges to that of the seed data, after which the pipeline is applied to the full dataset. Experiments on seven benchmarks show that DataEvolver significantly improves data quality, enabling LLMs to achieve an average improvement of 12\% and 3\% compared to using raw data and data produced by strong data preparation systems, respectively.

\section{Related Work}
\label{sec:rw}

\begin{figure*}[t]
	\centering
    \includegraphics[width=1\textwidth]{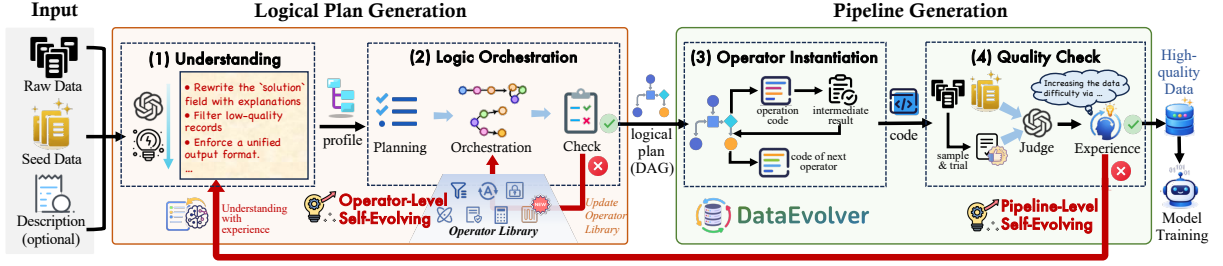}
    \caption{Architecture of DataEvolver. Given raw data and high-quality seed data (and optionally a task description), DataEvolver (1) understands the data to generate a data profile; (2) generates an executable logical plan via operator-level self-evolving; (3) instantiates the plan into code; and (4) performs quality checks to extract experience via pipeline-level self-evolving, ultimately transforming the raw data into high-quality data comparable to the seed data.}
	\label{fig:overview}
\end{figure*}

\subsection{Data Preparation for LLMs}

\textbf{Operator-based Systems.}
\blue{A major line of work develops operator libraries and recipe-style pipelines for scalable data cleaning, restructuring, and management. Data-Juicer~\cite{chen2023datajuiceronestopdataprocessing} formalizes data recipes with evaluation-aware workflows, and Data-Juicer 2.0~\cite{chen2025datajuicer20cloudscaleadaptive} extends this approach to cloud-scale and multimodal settings. DataTrove~\cite{penedo2024fineweb} supports scalable text processing, filtering, and deduplication, while Dolma~\cite{soldaini2024dolma} provides an open corpus and data curation tools for language model pretraining. These systems establish reusable infrastructure for data preparation.}
% To handle the scale of LLM training, recent engineering frameworks have established modular operator libraries and recipe-based execution models. These systems enable efficient, platform-agnostic processing of billions of tokens by formalizing data recipes and supporting cloud-scale multimodal workflows~\citep{chen2023datajuiceronestopdataprocessing, chen2025datajuicer20cloudscaleadaptive, penedo2024fineweb, soldaini2024dolma, liang2025dataflowllmdrivenframeworkunified}. While these infrastructures provide robust execution environments and evaluation-aware templates, the construction of the pipeline itself remains largely manual, relying on static, heuristic-driven configurations that lack adaptability to dynamic requirements.

\textbf{Pipeline Synthesis.}
Another line of work treats pipeline construction as synthesis or search, generating and refining workflows from task requirements. Text-to-pipeline~\cite{ge2025texttopipelinebridgingnaturallanguage} translates natural language into executable pipelines, while Prompt2DAG~\cite{alidu2025prompt2dagmodularmethodologyllmbased} assembles DAG-style workflows in a modular manner. LLaPipe~\cite{chang2025llapipellmguidedreinforcementlearning} explores operator spaces with LLM-guided reinforcement learning, and mlidea~\cite{10.14778/3750601.3750671} improves data preparation code via interactive diagnosis. These methods are primarily designed for tabular or traditional ML settings and provide limited support for seed-based target specification and stable alignment of LLM training data.
% Complementing static execution, another line of research treats pipeline construction as a program synthesis or search problem. By leveraging LLMs to interpret natural language requirements, these methods automatically generate executable DAG-style workflows or tabular processing scripts~\citep{ge2025texttopipelinebridgingnaturallanguage, alidu2025prompt2dagmodularmethodologyllmbased}. More advanced approaches incorporate reinforcement learning to explore operator spaces or utilize interactive diagnosis mechanisms to refine code quality~\citep{chang2025llapipellmguidedreinforcementlearning, 10.14778/3750601.3750671}. However, these synthesis methods are predominantly designed for structured data or traditional machine learning tasks, often lacking the specialized alignment mechanisms required for unstructured pretraining data.

\blue{Data selection and reweighting offer complementary ways to improve training corpora. DSIR~\cite{xie2023data} uses importance resampling to match a target distribution, while DoReMi~\cite{xie2023doremi} learns domain mixture weights through proxy-model optimization. DoPAMine~\cite{arannil2024dopaminedomainspecificpretrainingadaptation} combines seed-guided retrieval and classifier-based filtering for domain adaptation. DataEvolver uses seed-derived requirements to guide transformation pipelines that construct the desired supervision.}
%rather than data preparation.

\subsection{Self-Evolving Paradigm}
\blue{Feedback-driven refinement can improve generated outputs~\citep{madaan2023self} and reusable agent skills~\citep{wang2023voyager}. AgentEvolver~\citep{zhai2025agentevolverefficientselfevolvingagent}, LANCE~\citep{wang2025languagemodelscontinuousselfevolving}, and symbolic learning~\citep{zhou2024symboliclearningenablesselfevolving} explore related forms of self-evolution, as summarized in recent surveys~\citep{fang2025comprehensivesurveyselfevolvingai}. Data preparation increasingly incorporates related forms of agentic adaptation.}

% The self-evolving paradigm extends beyond static optimization, focusing on agents capable of iteratively updating their own code, tools, or memory structures based on environmental feedback. Representative works in embodied agents and software engineering demonstrate that LLMs can continuously refine their skills and logical processes through curriculum learning and symbol-based evolution~\citep{wang2023voyager, hong2023metagpt, zhai2025agentevolverefficientselfevolvingagent}. Within data-centric contexts, recent frameworks have begun to explore continuous self-refinement and symbolic learning to adapt to changing objectives~\citep{wang2025languagemodelscontinuousselfevolving, zhou2024symboliclearningenablesselfevolving, fang2025comprehensivesurveyselfevolvingai}. Despite this progress, there remains a gap in applying seed-anchored, self-evolving principles specifically to the optimization of multi-operator data preparation pipelines.

\blue{DataFlow-Agent supports operator synthesis and iterative execution verification~\cite{liang2025dataflowllmdrivenframeworkunified}, while DataFlow-Harness grounds DAG construction in a live registry and validates schema compatibility~\cite{he2026dataflowharness}. DataFoundry evolves preparation runtimes through pilot-set assessment and component adapters~\cite{yang2026datafoundry}; CurateEvo revises curation strategies using downstream agent development failures~\cite{wang2026curateevo}. DataEvolver integrates exemplar-guided target specification with two levels of self-evolution: (i) operator-level adaptation and logical-plan refinement, and (ii) pipeline-level trial-output assessment and experience-guided refinement. Coupling these loops connects structural refinement with feedback on prepared data before full-dataset execution. Appendix~\ref{sec:appendix:recent_comparison} provides further mechanism comparisons.}

% In summary, predefined systems~\cite{chen2023datajuiceronestopdataprocessing,chen2025datajuicer20cloudscaleadaptive,liang2025dataflowllmdrivenframeworkunified} emphasize scalable infrastructure but rely on manual recipe design. Pipeline synthesis and search~\cite{ge2025texttopipelinebridgingnaturallanguage,alidu2025prompt2dagmodularmethodologyllmbased,chang2025llapipellmguidedreinforcementlearning,10.14778/3750601.3750671} improve workflow construction but do not directly address seed-anchored, stable alignment for LLM training data. Seed-guided and self-evolving paradigms~\cite{arannil2024dopaminedomainspecificpretrainingadaptation,wang2025languagemodelscontinuousselfevolving,zhai2025agentevolverefficientselfevolvingagent,zhou2024symboliclearningenablesselfevolving,fang2025comprehensivesurveyselfevolvingai} provide complementary ideas but typically evolve different objects. Our work builds on these directions by evolving data preparation pipelines under seed guidance at multiple granularities to improve both executability and alignment to a target specification.

\section{Method}
\label{sec:method}

In this paper, we propose DataEvolver, which constructs an \emph{executable} and \emph{effective} data preparation pipeline through multi-level self-evolving, thereby automatically transforming raw data into high-quality data. In this section, we first formalize the problem and then introduce seed-guided data understanding, as well as operator-level and pipeline-level self-evolution mechanisms.

\subsection{Problem Formulation}
\label{sec:problem}

Given a raw dataset $\mathcal{D}_{raw} = \{x_i\}_{i=1}^{n}$, a set $\mathcal{S}$ of high-quality seed examples and an optional natural language description $\mathcal{T}$, our objective is to construct an effective data preparation pipeline $P^*$ from an operator library $\mathcal{O}$. Each logical operator specifies a transformation to be instantiated as executable code.

We represent a pipeline $P$ as a directed acyclic graph (DAG) of $k$ operators $\{o_1, o_2, \dots, o_k\}$, where each $o_j \in \mathcal{O}$ is parameterized by $\theta_j$. Writing $\Theta = \{\theta_j\}_{j=1}^{k}$, the data preparation process is:
\begin{gather}
\mathcal{D}_{out} = \mathrm{Exec}(P, \Theta, \mathcal{D}_{raw}),
\end{gather}
where $\mathrm{Exec}(\cdot)$ applies the operators in a topological order while respecting their data dependencies. DataEvolver refines $P^*$ to reduce discrepancies between the prepared data $\mathcal{D}_{out}$ and the quality requirements $\mathcal{C}$ derived from $\mathcal{S}$. The effectiveness of $\mathcal{D}_{out}$ is further evaluated through downstream LLM training or proxy evaluation.

\subsection{Seed-Guided Data Understanding}
\label{sec:understanding}

Understanding the characteristics of high-quality data is central to constructing an effective preparation pipeline. DataEvolver uses a small seed set $\mathcal{S}$ to abstract the desired output structure, formatting, and quality requirements. Combined with the raw-data characteristics and accumulated experience, this understanding provides a structured specification for operator orchestration and pipeline refinement.

Specifically, DataEvolver proposes seed-guided data understanding, which extracts a structured data profile $\mathcal{C}$ based on some sampled raw data $\mathcal{D}_{sample}$, seed data $\mathcal{S}$, and evolutionary experience $\mathcal{M}$:
\begin{gather}
\mathcal{C} 
= f_{\text{understand}}\bigl(\mathcal{S}, \mathcal{D}_{sample}, \mathcal{M}\bigr)  \label{eq2}
\end{gather}
where $f_{\text{understand}}$ denotes the profile induction function. $\mathcal{M}$ is the memory of cumulative experience. $\mathcal{M}$ is initialized with the user-provided natural language description $\mathcal{T}$ (if no input is provided, $\mathcal{M}$ is empty), and is continuously updated with new evolutionary experience during pipeline-level self-evolving.

% In our framework, the seed set $\mathcal{S}$ serves as a lightweight specification of the target data distribution rather than a collection of demonstrations to be copied verbatim. Each seed is curated to reflect common best practices for high-quality SFT data, such as clear task intent, well-formed output structure, task-faithful formatting, and normalized answers. Collectively, the seeds expose recurring patterns and critical requirements of the target data, including required fields, structural organization, stylistic preferences, and quality constraints. This allows DataEvolver to infer not only what fields should be produced, but also how the final supervision should be organized and expressed.

% p9
% In our framework, the seed set $S$ serves as a lightweight specification of the target data rather than a collection of demonstrations to be copied verbatim. Instead of explicitly defining the whole pipeline, the seeds reveal key characteristics of the desired outputs, such as structure, formatting, normalization, and quality expectations. DataEvolver then abstracts these recurring characteristics into a structured data profile $C$ for pipeline construction.
In our framework, the seed set $\mathcal{S}$ is used as a lightweight specification of the target data, from which DataEvolver abstracts recurring structural, formatting, and quality requirements into the data profile $\mathcal{C}$. In practice, $\mathcal{C}$ summarizes the target output schema, formatting conventions, quality constraints, and difficulty preferences. A concrete example is shown in the \textit{Understanding Result} block of Figure~\ref{fig:case_1} in Appendix~\ref{sec:appendix:case_study}.

%In practice, $\mathcal{C}$ captures requirements over structure and fields, formatting and style, quality constraints and difficulty preferences. Appendix \ref{sec:appendix:case_study} shows an example of data profile $\mathcal{C}$.

\subsection{Operator-Level Self-Evolving}
\label{sec:op-evolve}

Constructing a long-chain data processing pipeline is inherently prone to logical failures \cite{jiang2023selfplanning}. To mitigate these risks, DataEvolver begins with operator-level self-evolving, aiming to establish a \emph{logically executable} plan prior to any physical execution.

We model the logical plan as a DAG $G = (V, E)$, where nodes $V$ represent operators from $\mathcal{O}$ and edges $E$ capture data dependencies. Logical checks identify dependency gaps, interface incompatibilities, and structural flaws (e.g., disconnected components). The system repairs these issues by inserting bridging operators, substituting incompatible ones, reordering nodes, or rewiring edges~\cite{chen2023selfdebug}. When the available operators lack a required capability, DataEvolver generates new operators tailored to the gap, incorporates them into $\mathcal{O}$, and adjusts the DAG. This iterative refinement adapts both the operator library and the plan before code instantiation.

Formally, the operator-level self-evolving can be expressed as:
\begin{gather}
G^{(t+1)}, \mathcal{O}^{(t+1)} = f_{\text{op-evolving}}\bigl(G^{(t)}, \mathcal{O}^{(t)}, \mathcal{C}\bigr),
% \\ \mathcal{O}^{(t+1)} = \mathcal{O}^{(t)} \cup \{o_{\text{new}}\},
\end{gather}
where $f_{\text{op-evolving}}$ denotes the evolving function (repair and re-generation). $G^{(t)}$ and $\mathcal{O}^{(t)}$ denote the DAG and operator library at iteration $t$. $\mathcal{O}^{(t+1)} = \mathcal{O}^{(t)} \cup \{o_{\text{new}}\}$ denotes the augmented operator library after generating a new operator $o_{\text{new}}$ in order to improve $G^{(t)}$. Through iteration, DataEvolver produces a logical plan of operators.

\subsection{Pipeline-Level Self-Evolving}
\label{sec:pipe-evolve}

Operator-level self-evolving refines logical operators and their dependencies before code instantiation. Pipeline-level self-evolving complements these checks with feedback from trial execution, revealing how the instantiated pipeline transforms actual data. It compares trial outputs with seed-implied quality requirements and summarizes discrepancies as experience for refining the preparation pipeline.

Starting from the logical plan \( G^{*} \), DataEvolver instantiates each operator into code with parameters \( \Theta = \{\theta_j\} \) to form a concrete pipeline \( P \). Rather than immediately applying \( P \) to the full dataset, the system conducts trial runs on a small sampled subset \( \mathcal{D}_{s} \subset \mathcal{D}_{raw} \), producing trial outputs $\mathcal{D}'_{s} = P(\mathcal{D}_{s})$.
These outputs are then compared with the seed data \( \mathcal{S} \) to generate \emph{discrepancy signals}, which indicate whether certain aspects of data quality are under-satisfied, over-constrained, or missing. DataEvolver converts these signals into \emph{evolutionary experience} and accumulates them in an experience memory \( \mathcal{M} \).
Formally, pipeline-level self-evolving can be expressed as:
\begin{gather}
\begin{aligned}
\mathcal{M}^{(t+1)} &= f_{\text{pipe-evolving}}\bigl(\mathcal{M}^{(t)}, Exp \bigr),\\
\text{where}\; Exp &= f_{\text{judge}}\bigl(P^{(t)}(\mathcal{D}_{s}), \mathcal{S}\bigr),
\end{aligned}
\end{gather}
Here, \( \mathcal{M}^{(t)} \) is the accumulated experience and \( P^{(t)} \) the instantiated pipeline at iteration \( t \). The judge \( f_{\text{judge}} \) assesses trial outputs against the seed-implied quality requirements and generates experience \( Exp \) describing unmet requirements. The function \( f_{\text{pipe-evolving}} \) uses this feedback to update the memory to \( \mathcal{M}^{(t+1)} \).

With the updated memory \( \mathcal{M}^{(t+1)} \), DataEvolver recomputes the data profile \( \mathcal{C} \) and re-orchestrates the logical plan through operator-level self-evolving before instantiating the next pipeline. Trial-output experience thus guides both understanding and orchestration, coupling the two evolving levels. Once the trial outputs meet the seed-implied quality requirements, the refined pipeline is applied to the full dataset.
%Core prompts used in these stages are provided in Appendix~\ref{sec:appendix:prompts}.

\section{Experiments}
\label{sec:exp}

\begin{table*}[t]
\centering
\small
\setlength{\tabcolsep}{3.5pt}
\begin{tabular}{llcccccccc}
\toprule
& & & \multicolumn{1}{c}{\textbf{General}} & \multicolumn{2}{c}{\textbf{QA--MCQ}} & \multicolumn{2}{c}{\textbf{Math Reasoning}} & \multicolumn{2}{c}{\textbf{Text-to-SQL}} \\
\cmidrule(lr){4-4}\cmidrule(lr){5-6}\cmidrule(lr){7-8}\cmidrule(lr){9-10}
\textbf{LLM Backbones} & \textbf{Training Data} & \textbf{Size} &
\textbf{Alpaca} & \textbf{ARC-E} & \textbf{ARC-C} & \textbf{GSM8K} & \textbf{MATH} & \textbf{Spider} & \textbf{BIRD} \\
& & &
Win-rate & Acc. & Acc. & EM & EM & Exec-Acc & Exec-Acc \\
\midrule

\multirow{7}{*}{\textbf{Qwen3-8B-Base}}
& Base Model& -- & 45.3 & 65.43 & 59.66 & 83.17 & 55.98 & 45.16 & 20.80 \\ \cmidrule{2-10}
& Vanilla SFT & 1k & 42.7 & 53.12 & 49.92 & 85.23 & 48.40 & 50.74 & 42.82 \\
& Vanilla SFT & 5k & 44.8 & 55.74 & 51.25 & 85.75 & 49.68 & 52.10 & 41.45 \\\cmidrule{2-10}
& DataFlow-SFT & 1k & 52.6 & 68.11 & 63.84 & 86.28 & 68.68 & 55.83 & 45.70 \\
& DataFlow-SFT & 5k & 51.9 & 68.83 & 64.93 & 87.16 & \textbf{69.28} & 55.26 & 46.33 \\\cmidrule{2-10}
& DataEvolver-SFT & 1k & 55.0 & 71.56 & \textbf{66.73} & 86.92 & 67.36 & 59.04 & 50.08 \\
& DataEvolver-SFT & 5k & \textbf{56.3} & \textbf{73.37} & 66.09 & \textbf{87.83} & 68.42 & \textbf{60.43} & \textbf{50.34} \\
\midrule

\multirow{7}{*}{\textbf{Gemma-3-4B-it}}
& Base Model& -- & 48.6 & 82.19 & 72.54 & 84.91 & 61.44 & 65.38 & 32.86 \\\cmidrule{2-10}
& Vanilla SFT & 1k & 41.2 & 88.24 & 78.91 & 86.21 & 55.06 & 68.36 & 38.46 \\
& Vanilla SFT & 5k & 42.0 & 89.03 & 77.53 & 85.13 & 56.85 & 69.12 & 39.01 \\\cmidrule{2-10}
& DataFlow-SFT & 1k & 57.8 & 95.76 & 82.45 & 88.56 & 64.78 & 75.86 & 48.93 \\
& DataFlow-SFT & 5k & 58.3 & 96.25 & 83.21 & 89.63 & 64.21 & 74.93 & 50.77 \\\cmidrule{2-10}
& DataEvolver-SFT & 1k & \textbf{61.4} & 95.48 & 85.36 & 90.45 & 66.27 & 75.28 & 52.35 \\
& DataEvolver-SFT & 5k & 61.1 & \textbf{97.83} & \textbf{86.67} & \textbf{90.98} & \textbf{67.58} & \textbf{77.02} & \textbf{53.62} \\
\midrule

\multirow{7}{*}{\textbf{Llama3.1-8B-Instruct}}
& Base Model& -- & 25.9 & 91.18 & 82.37 & 84.38 & 42.76 & 63.25 & 38.07 \\\cmidrule{2-10}
& Vanilla SFT & 1k & 39.1 & 85.72 & 85.23 & 86.80 & 50.76 & 60.14 & 44.21 \\
& Vanilla SFT & 5k & 39.4 & 86.19 & 87.21 & 89.43 & 52.33 & 61.92 & 43.18 \\\cmidrule{2-10}
& DataFlow-SFT & 1k & 46.6 & 94.26 & 90.10 & 88.21 & 63.91 & 70.48 & 49.36 \\
& DataFlow-SFT & 5k & 48.0 & 96.03 & 90.88 & 88.53 & 64.58 & 71.39 & 49.05 \\\cmidrule{2-10}
& DataEvolver-SFT & 1k & 51.7 & 96.30 & 91.75 & \textbf{89.72} & 65.62 & 70.15 & \textbf{53.77} \\
& DataEvolver-SFT & 5k & \textbf{52.5} & \textbf{98.47} & \textbf{92.03} & 89.34 & \textbf{67.74} & \textbf{72.66} & 52.85 \\
\bottomrule
\end{tabular}
\caption{Main results of SFT with training data prepared by different methods. We compare Vanilla SFT, DataFlow-SFT, and DataEvolver-SFT. The best results are highlighted in \textbf{bold}.}
\label{tab:main_exp}
\end{table*}

\subsection{Benchmarks and Baselines}
\label{sec:exp:bench_baseline}
To evaluate the quality of data preparation systems, we consider the performance of LLMs trained on data prepared by different systems. We conduct experiments on 7 benchmarks from four categories: instruction following (Alpaca; AlpacaEval~2.0 win-rate)~\cite{dubois2024lc_alpacaeval}, multiple-choice QA (ARC-Easy/Challenge; accuracy)~\cite{clark2018arc}, math reasoning (GSM8K/MATH; exact match)~\cite{cobbe2021trainingverifiers,hendrycks2021math}, and Text-to-SQL (Spider/BIRD; execution accuracy)~\cite{yu2018spider,li2023bird}. Experiments are conducted on various LLM backbones, including \texttt{Qwen3-8B-Base}~\cite{yang2025qwen3}, \texttt{Gemma-3-4B-it}~\cite{gemma2025gemma3}, and \texttt{Llama3.1-8B-Instruct}~\cite{llama3}. We compare Vanilla SFT (using the original training data), DataFlow-SFT~\cite{liang2025dataflowllmdrivenframeworkunified} (a strong LLM-oriented data preparation system), and the proposed DataEvolver-SFT.

\subsection{Experimental Setup}
\label{sec:exp:setup}

\textbf{Setting.}
We evaluate the data preparation system under the most common SFT setting. \blue{Raw data are drawn from the official training sources, and the reported 1k/5k budgets denote final SFT instances. For math reasoning, GSM8K and MATH form a shared raw pool, with evaluation reported separately on each benchmark.} Training and evaluation are conducted using \texttt{ms-swift}~\cite{modelscope2024msswift} and \texttt{OpenCompass}~\cite{opencompass2023}, respectively. All methods use the same training and evaluation protocols.

\textbf{Seed Specification.}
DataEvolver takes 20 seed data examples as a lightweight specification. \blue{The seed examples specify the desired supervision format and quality criteria.} In practice, seed data can come from various sources, such as previously curated data or human annotations. DataEvolver is effective with only a small number of seed examples, avoiding excessive manual effort. Appendix~\ref{sec:appendix:seed_data} shows some examples of seed data, and Appendix~\ref{sec:appendix:seed_robustness} provides additional analyses on seed robustness.

\subsection{Main Results}
\label{sec:exp:main}

\textbf{LLM Performance.}
To evaluate whether DataEvolver improves SFT performance, we compare Vanilla SFT, DataFlow-SFT, and DataEvolver-SFT across seven benchmarks using multiple LLM backbones. \blue{As shown in Table~\ref{tab:main_exp}, DataEvolver-SFT achieves an average improvement of approximately 10 score points over Vanilla SFT across the 42 model--budget--benchmark settings, demonstrating substantial data quality enhancement. Compared with DataFlow-SFT, DataEvolver yields an average improvement of approximately 2 score points. These are unweighted descriptive averages of benchmark scores on their reported 0--100 scales. DataEvolver uses seed-derived specifications and complementary feedback at the operator and pipeline levels to refine the preparation workflow.}

Moreover, we find that using 1k data prepared by DataEvolver achieves performance comparable to using 5k data. This indicates that DataEvolver significantly enhances data quality, aligning with previous findings that a small amount of high-quality data can lead to strong performance \citep{chen2024alpagasustrainingbetteralpaca,du2023modsmodelorienteddataselection,lu2023instaginstructiontagginganalyzing,zhou2023limaalignment}. \blue{The results across four task groups further demonstrate the flexibility of a common preparation framework: DataEvolver adapts its operators and pipeline to the raw data and seed specification. Overall, its multi-level self-evolution provides a unified approach to automatic data preparation for LLMs.}

\begin{figure*}[t]
  \centering
  \includegraphics[width=\textwidth]{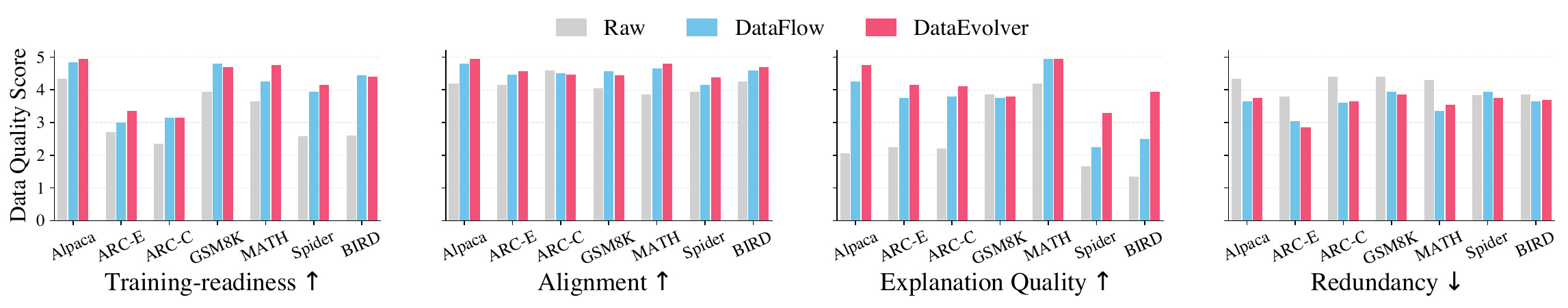}
  \caption{Data quality evaluation results for training-readiness, seed alignment, explanation quality, and redundancy.}
  \label{fig:llm_judge_score}
\end{figure*}

% \begin{table}[t]
% \centering
% \footnotesize
% % \setlength{\tabcolsep}{3.0pt}
% % \renewcommand{\arraystretch}{1.08}
% \resizebox{\columnwidth}{!}{%
% \begin{tabular}{@{}llccr@{}}
% \toprule
% \textbf{Tasks} & \textbf{Datasets} &
% \multicolumn{1}{c}{\textbf{DataFlow}} &
% \multicolumn{1}{c}{\textbf{DataEvolver}} &
% \multicolumn{1}{c}{\textbf{$\Delta$}} \\
% & &
% \multicolumn{1}{c}{\textbf{(Tok/item)}} &
% \multicolumn{1}{c}{\textbf{(Tok/item)}} &
%  \\
% \midrule
% \multirow{1}{*}{General-Inst.} & Alpaca & 3,214 & 1,225 & -61.9\% \\
% \addlinespace[1pt]
% \multirow{2}{*}{QA--MCQ} & ARC-Easy & 3,768 & 1,652 & -56.2\% \\
% & ARC-Challenge & 4,091 & 2,135 & -47.8\% \\
% \addlinespace[1pt]
% \multirow{2}{*}{Math Reasoning} & GSM8K & 2,564 & 1,866 & -27.2\% \\
% & MATH & 4,832 & 2,960 & -38.7\% \\
% \addlinespace[1pt]
% \multirow{2}{*}{Text-to-SQL} & Spider & 2,592 & 1,926 & -25.7\% \\
% & BIRD & 2,755 & 2,118 & -23.1\% \\
% \midrule
% \multicolumn{2}{l}{\textbf{Macro avg (7 datasets)}} &
% \textbf{3,402} & \textbf{1,983} & \textbf{-40.1\%} \\
% \bottomrule
% \end{tabular}%
% }
% \caption{Token cost comparison between DataFlow and DataEvolver during data preparation. Tok/item is the average tokens per generated training item during data preparation (lower is better).}

% \label{tab:token_cost}
% \end{table}

\begin{table}[t]
\centering\small
\begin{tabular}{llC{1cm}C{1cm}c}
\toprule
\textbf{Tasks}                  & \textbf{Datasets} & \textbf{\begin{tabular}[c]{@{}c@{}}$\!\!\!\!$DataFlow\\ $\!\!\!\!$(Tok/item)\end{tabular}} & \textbf{\begin{tabular}[c]{@{}c@{}}$\!\!\!$DataEvolver\\ $\!\!\!$(Tok/item)\end{tabular}} & \multicolumn{1}{c}{\textbf{$\Delta$}} \\ \midrule
General                   & Alpaca            & 3,214                                                                  & 1,225                                                                     & -61.9\%                               \\
\multirow{2}{*}{QA-MCQ}        & ARC-E          & 3,768                                                                  & 1,652                                                                     & -56.2\%                               \\
                                & ARC-C     & 4,091                                                                  & 2,135                                                                     & -47.8\%                               \\
\multirow{2}{*}{Math} & GSM8K             & 2,564                                                                  & 1,866                                                                     & -27.2\%                               \\
                                & MATH              & 4,832                                                                  & 2,960                                                                     & -38.7\%                               \\
\multirow{2}{*}{Text-to-SQL}    & Spider            & 2,592                                                                  & 1,926                                                                     & -25.7\%                               \\
                                & BIRD              & 2,755                                                                  & 2,118                                                                     & -23.1\%                               \\ \midrule
\multicolumn{2}{c}{\textbf{Macro Avg.}}             & \textbf{3,402}                                                         & \textbf{1,983}                                                            & \textbf{-40.1\%}                      \\ \bottomrule
\end{tabular}
\caption{Token cost of DataEvolver during data preparation. Tok/item is the average tokens per generated training item (lower is better). More statistics on data preparation overhead are reported in Appendix \ref{sec:appendix:pipeline_stats}.}
\label{tab:token_cost}
\end{table}

\textbf{Data Quality Evaluation.}
To directly assess the quality of the prepared data, following \citet{liang2025dataflowllmdrivenframeworkunified}, we employ an LLM-based judge to score the data on a 1–5 scale along four dimensions: training-readiness, seed alignment, explanation quality, and redundancy. As shown in Figure~\ref{fig:llm_judge_score}, DataEvolver improves both training-readiness and seed alignment over raw data and the data prepared by DataFlow, with the largest gains typically observed in explanation quality. 

Meanwhile, DataEvolver also reduces redundancy (lower is better), reflecting less repetitive fields or content in the prepared instances \cite{abbas2023semdedup}. Notably, the data processed by DataEvolver exhibits lower redundancy while simultaneously improving downstream model performance after training, further demonstrating its positive impact on data quality. 
Overall, DataEvolver achieves higher average training-readiness, seed alignment, and explanation quality than DataFlow, while producing less redundant training data. \blue{The existing two-annotator comparison across four task groups yields human--judge Spearman correlations of 0.84--0.91 (Appendix~\ref{sec:appendix:judge_validation}).}

\textbf{Token Cost.}
In the data preparation pipeline, some operators may rely on invoking advanced LLMs, which leads to token consumption during pipeline execution. To compare data preparation efficiency, we measure the LLM token consumption incurred during data preparation for DataEvolver and DataFlow \cite{liang2025dataflowllmdrivenframeworkunified}. As shown in Table~\ref{tab:token_cost}, DataEvolver consistently reduces the amortized token cost per generated training instance across all datasets, \blue{achieving a 40.1\% macro-average reduction, computed as the mean of per-dataset percentage reductions.} Overall, DataEvolver not only improves LLM performance but also attains substantially lower token cost during data preparation. Further results are reported in Appendix~\ref{sec:appendix:wall_time} and ~\ref{sec:appendix:token_matched}.
%Detailed pipeline statistics are provided in Appendix \ref{sec:appendix:pipeline_stats}.
Detailed pipeline statistics and synthesized operator examples are provided in Appendix~\ref{sec:appendix:pipeline_stats}.

% p9
% \textbf{Operator Synthesis.}
% Beyond improving effectiveness and efficiency, DataEvolver can also automatically extend its operator inventory when predefined operators are insufficient for a target task. In our experiments, DataEvolver starts from a shared set of task-agnostic operators for generic data transformation, and synthesizes only a small number of task-specific operators when capability gaps are detected during operator-level self-evolving. These synthesized operators are tightly coupled to the requirements implied by the seeds: for General instruction and QA-MCQ tasks, they mainly enforce standardized instance construction and answer/format consistency; for Math Reasoning, they support normalization, step decomposition, and final-answer verification; for Text-to-SQL, they focus on schema grounding and evidence fusion. This behavior shows that DataEvolver does not rely on manually predefined task-specific recipes, but can minimally expand a shared operator library to better match heterogeneous task requirements. More detailed operator examples are provided in Appendix~\ref{sec:appendix:pipeline_stats}.

\section{Analysis}
\label{sec:analysis}

We conduct an in-depth study of DataEvolver, with additional analyses provided in the appendix.

\subsection{Ablation of Multi-level Self-Evolving}
\label{sec:analysis:ablation}

\begin{table*}[t]
\centering
\small
\begin{tabular}{lccccccc}
\toprule
\multirow{3}{*}{\textbf{Methods}} &
\multicolumn{1}{c}{\textbf{General}} &
\multicolumn{2}{c}{\textbf{QA--MCQ}} &
\multicolumn{2}{c}{\textbf{Math Reasoning}} &
\multicolumn{2}{c}{\textbf{Text-to-SQL}} \\
\cmidrule(lr){2-2}\cmidrule(lr){3-4}\cmidrule(lr){5-6}\cmidrule(lr){7-8}
& \makecell[c]{\textbf{Alpaca}\\Win-rate} &
\makecell[c]{\textbf{ARC-E}\\Acc.} &
\makecell[c]{\textbf{ARC-C}\\Acc.} &
\makecell[c]{\textbf{GSM8K}\\EM} &
\makecell[c]{\textbf{MATH}\\EM} &
\makecell[c]{\textbf{Spider}\\Exec-Acc} &
\makecell[c]{\textbf{BIRD}\\Exec-Acc} \\
\midrule
Base Model & 25.9 & 91.18 & 82.37 & 84.38 & 42.76 & 63.25 & 38.07 \\
\midrule
DataEvolver                  & \textbf{52.5} & \textbf{98.47} & \textbf{92.03} & \textbf{89.34} & \textbf{67.74} & \textbf{72.66} & \textbf{52.85} \\
$\;\;\;$w/o operator-level self-evolving   & 42.8 & 86.30 & 86.75 & 87.02 & 61.93 & 70.09 & 47.46\\
$\;\;\;$w/o pipeline-level self-evolving & 48.7 & 94.09 & 90.27 & 86.11 & 62.08 & 66.53 & 45.63 \\
\bottomrule
\end{tabular}
\caption{Ablation study of the two self-evolving components on \texttt{Llama3.1-8B-Instruct}. We compare the base model, two variants (removing operator-level or pipeline-level self-evolving), and DataEvolver.}

\label{tab:ablation_exp}
\end{table*}

DataEvolver adopts a multi-level self-evolving mechanism that integrates operator-level self-evolving and pipeline-level self-evolving. To verify whether the performance gains of DataEvolver truly stem from the collaboration of these two components, we conduct an ablation study by selectively disabling each self-evolving loop in Table~\ref{tab:ablation_exp}.

The results show that removing either component consistently degrades performance across benchmarks, while the full system achieves the strongest overall results. The degradation is particularly evident on tasks that rely on well-structured intermediate transformations and iterative quality refinement. This indicates that operator-level self-evolving is crucial for constructing a logically sound and executable pipeline, whereas pipeline-level self-evolving further refines data quality through execution feedback.

\subsection{Effect of Operator-level Self-Evolving}
\label{sec:analysis:op_level}

\textbf{Operator-level Self-evolving Improves Logical Plan Quality.}
Figure~\ref{fig:op_level_analysis} in Appendix~\ref{sec:appendix:plan_assessment} evaluates operator logical plans with and without the operator-level self-evolving loop (via LLM-as-a-judge \cite{zheng2023judging}). Enabling this loop consistently enhances logical plan quality across three complementary dimensions: (i) \emph{operator adequacy}, indicating fewer missing capabilities required to achieve the target I/O transformation; (ii) \emph{interface consistency}, reflecting improved alignment of intermediate field names and types between adjacent operators; and (iii) \emph{plan coherence}, suggesting fewer structural inconsistencies in the composed plan. Overall, these improvements demonstrate that operator-level refinement yields plans that are both more complete and more structurally compatible, laying a solid foundation for subsequent code instantiation.

\textbf{DataEvolver Synthesizes Operators Automatically.}
Furthermore, when required operators are missing, DataEvolver can synthesize new operators during the operator-level self-evolving process and incorporate them into the operator library. This capability allows DataEvolver to flexibly adapt to data preparation tasks across different domains. As shown in Appendix \ref{sec:appendix:op_details}, beyond a set of common foundational operators, DataEvolver autonomously generates domain-specific specialized operators when handling data from different domains, thereby improving data quality. In this way, operator synthesis remains minimal and targeted, while still improving structural compatibility and execution stability across heterogeneous tasks. Collectively, this enables DataEvolver to function as an evolving data preparation system.

% \subsection{Effect of Iterative Refinement}
\subsection{Effect of Pipeline-level Self-Evolving}
\label{sec:analysis:iter}

To verify that pipeline-level self-evolving leads to progressive improvements, we evaluate DataEvolver across multiple iterations: the initial setting uses the raw training split for SFT, and later iterations reuse the experience collected from previous runs to refine the understanding stage and pipeline construction.

\begin{figure*}[t]
  \centering
  \begin{subfigure}[t]{0.245\textwidth}
    \centering
    \includegraphics[width=\linewidth]{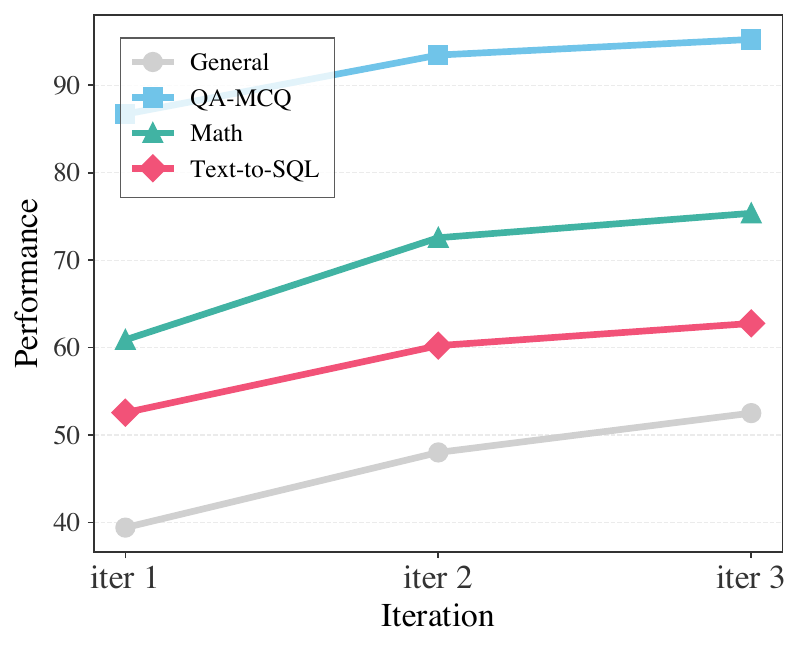}    \captionsetup{width=0.9\linewidth}
    \caption{LLM performance trained on the prepared data.}
    \label{fig:iteration_analysis:a}
  \end{subfigure}
  \hfill
  \begin{subfigure}[t]{0.245\textwidth}
    \centering
    \includegraphics[width=\linewidth]{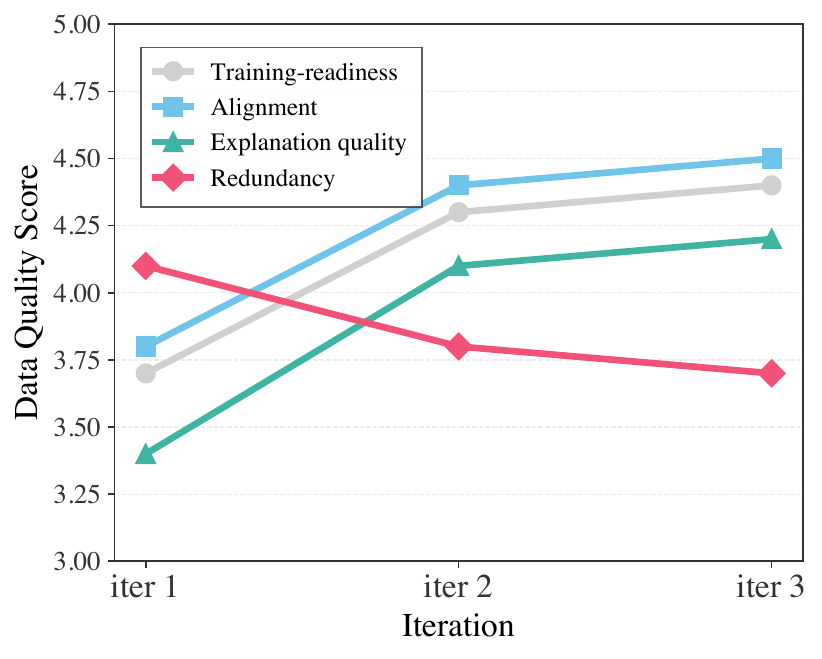}
    \captionsetup{width=0.9\linewidth}
    \caption{Data quality scores of the prepared data.}
    \label{fig:iteration_analysis:b}
  \end{subfigure}
  \hfill
  \begin{subfigure}[t]{0.245\textwidth}
    \centering
    \includegraphics[width=\linewidth]{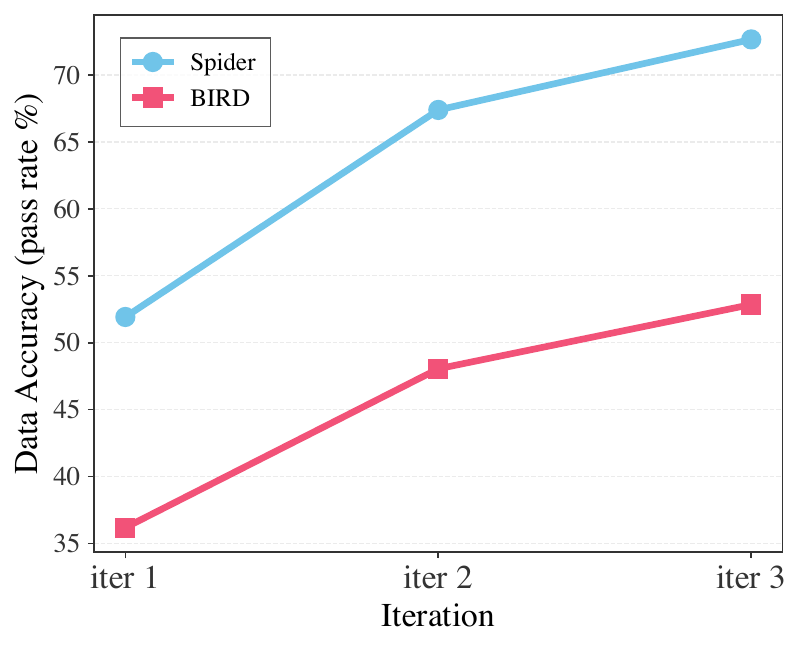}
    \captionsetup{width=0.95\linewidth}
    \caption{Data accuracy (whether the prepared SQL is correct).}
    \label{fig:iteration_analysis:c}
  \end{subfigure}
  \hfill
  \begin{subfigure}[t]{0.245\textwidth}
    \centering
    \includegraphics[width=\linewidth]{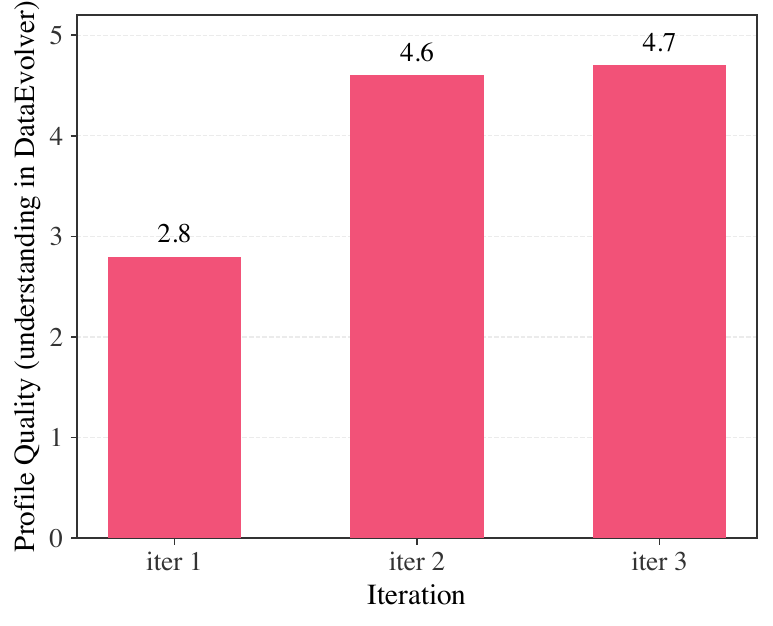}
    \captionsetup{width=0.95\linewidth}
    \caption{Profile quality during data preparation (i.e., Eq. (\ref{eq2})).}
  \label{fig:seed_understanding}
  \end{subfigure}

  % \caption{Iteration effects of pipeline-level self-evolving on \texttt{Llama3.1-8B-Instruct}. `iter-0' uses raw data for SFT, `iter-1/2/3' use DataEvolver-prepared data after 1/2/3 iterations with experience feedback.}
  \caption{Iteration effects of pipeline-level self-evolving on \texttt{Llama3.1-8B-Instruct}.}
  \label{fig:iteration_analysis}
\end{figure*}

Figure~\ref{fig:iteration_analysis}(a/b/c) summarizes the effects on \texttt{Llama3.1-8B-Instruct}. Downstream performance improves across task groups, with the largest gain typically appearing after the first iteration. Data quality scores also improve: training readiness, seed alignment, and explanation quality increase, while redundancy decreases. For Text-to-SQL, the SQL pass rate consistently increases on both Spider and BIRD, showing that iterative evolving improves the syntactic and structural validity of generated SQL supervision. Overall, pipeline-level self-evolving progressively enhances data preparation effectiveness.

\subsection{Pipeline Executability Diagnostics}
\label{sec:analysis:exec}

\begin{table}[t]
\centering\small
\setlength{\tabcolsep}{3pt}
\resizebox{\columnwidth}{!}{%
\begin{tabular}{@{}lccc@{}}
\toprule
\textbf{Failure Types} & \textbf{DataEvolver} & \begin{tabular}[c]{@{}c@{}}\textbf{w/o}\\ \textbf{operator-level}\end{tabular} & \begin{tabular}[c]{@{}c@{}}\textbf{w/o}\\ \textbf{pipeline-level}\end{tabular} \\ \midrule
Missing dependency    & 0           & 15                                                           & 4                                                            \\
Interface mismatch    & 0           & 9                                                            & 4                                                            \\
DAG disconnected      & 0           & 5                                                            & 7                                                            \\
Ordering conflict     & 0           & 3                                                            & 2                                                            \\ \midrule
Total     & 0           & 32                                                           & 17                                                           \\ \bottomrule
\end{tabular}%
}
\caption{Counts of different failure types during pipeline generation on \textsc{MATH} across 10 runs.}
\label{tab:error_types}
\end{table}

% \begin{table}[]
% \footnotesize
% \begin{tabular}{lccc}
% \hline
% \textbf{Failure Type} & DataEvolver & \begin{tabular}[c]{@{}c@{}}w/o\\ operator-level\end{tabular} & \begin{tabular}[c]{@{}c@{}}w/o\\ pipeline-level\end{tabular} \\ \hline
% Missing dependency    & 0           & 15                                                           & 4                                                            \\
% Interface mismatch    & 0           & 9                                                            & 4                                                            \\
% DAG disconnected      & 0           & 5                                                            & 7                                                            \\
% Ordering conflict     & 0           & 3                                                            & 2                                                            \\ \hline
% Observed failures     & 0           & 32                                                           & 17                                                           \\ \hline
% \end{tabular}
% \end{table}

To disentangle how individual components contribute to pipeline executability, we analyze the failure types under two ablation settings in Table~\ref{tab:error_types}. The two self-evolving loops exhibit clearly distinct failure profiles, highlighting their complementary roles.
% To disentangle how individual components contribute to the executability of the data preparation pipeline, we analyze the types of pipeline failures. Table~\ref{tab:error_types} compares the distributions of failure types under two ablation settings. Two levels of self-evolving exhibit clearly distinct failure profiles, highlighting the complementary roles of the two self-evolving loops.

When operator-level self-evolving is removed, failures are dominated by \emph{missing dependencies} and \emph{interface mismatches}. This suggests that, under a fixed operator inventory and without localized repair, the pipeline often fails to generate required intermediate artifacts or satisfy downstream input--output contracts, causing errors to concentrate on step-level feasibility. In contrast, disabling pipeline-level self-evolving shifts failures toward \emph{global structural issues}, especially \emph{disconnected DAGs}, indicating that one-shot plan composition is brittle at the graph level~\cite{wu2023autogen, xi2023rise}. Even when individual operators are locally plausible, the overall wiring can remain invalid without discrepancy feedback from trial executions.
% When operator-level self-evolving is removed, failures are dominated by \emph{missing dependencies} and \emph{interface mismatches}. This pattern suggests that, under a fixed operator inventory and without localized repair or closure mechanisms, the pipeline frequently fails to generate required intermediate artifacts or to satisfy downstream input–output contracts, causing errors to concentrate on step-level feasibility \cite{parameswaran2020data}. In contrast, disabling pipeline-level self-evolving shifts failures toward \emph{global structural issues}, most prominently a substantially higher proportion of \emph{disconnected DAGs}. This indicates that one-shot plan composition is brittle at the graph level \cite{wu2023autogen, xi2023rise}: even when individual operators are locally plausible, the overall wiring and dependency structure can be invalid in the absence of discrepancy feedback from trial executions to revise the plan.

Overall, operator-level self-evolving primarily enhances \emph{local validity} through capability closure and interface compatibility, whereas pipeline-level self-evolving improves \emph{global validity} by enforcing graph connectivity and compositional correctness. Their complementary effects explain why integrating both loops yields the most reliable end-to-end performance.

% Overall, this diagnostic evidence supports our motivation for multi-level self-evolving. Operator-level self-evolving primarily enhances \emph{local validity} by ensuring capability closure and interface compatibility, whereas pipeline-level self-evolving improves \emph{global validity} by enforcing graph connectivity and compositional correctness. Their complementary effects explain why integrating both loops yields the most reliable end-to-end performance observed in downstream experiments.

% \subsection{Alignment Signals and Structural Comparison}
% \subsection{Superiority of Seed Data Understanding}
% \subsection{$\!\!$Effect of Seed-Guided Data Understanding}
\subsection{Effect of Seed-Guided Data Understanding}
\label{sec:analysis:seed}

We analyze how experience feedback improves seed-driven data understanding, which serves as the specification for pipeline construction. To this end, Figure~\ref{fig:iteration_analysis}(d) evaluates the quality of the data profile $\mathcal{C}$ at each iteration of pipeline-level self-evolving. The results show that the quality of the data profile consistently improves throughout the evolution process. Specifically, after each iteration, DataEvolver compares a small batch of prepared examples with the seed specification, summarizes recurring discrepancies (e.g., missing or mismatched fields and redundant content), and translates them into actionable constraints for the next round of refinement. 
% The progressively strengthened data profile enables DataEvolver to continuously improve the
The progressively strengthened data profile enables DataEvolver to continuously improve data preparation quality. In particular, the \textit{Understanding Result} block in Figure~\ref{fig:case_1} in Appendix~\ref{sec:appendix:case_study} illustrates a concrete instance of the induced data profile $C$.

% Existing Appendix limited-seed results, transposed for main-text visibility.
% The full original table and all its baseline rows remain in the appendix.
\begin{table}[ht]
\centering
\small
\blue{\begin{tabular*}{\columnwidth}{@{\extracolsep{\fill}}lrr@{}}
\toprule
\textbf{Benchmark} & \textbf{20 seeds} & \textbf{5 seeds} \\
\midrule
Alpaca & 52.5 & 51.4 \\
ARC-E & 98.47 & 98.02 \\
ARC-C & 92.03 & 90.24 \\
GSM8K & 89.34 & 89.02 \\
MATH & 67.74 & 64.98 \\
Spider & 72.66 & 72.45 \\
BIRD & 52.85 & 51.73 \\
\bottomrule
\end{tabular*}}
\caption{\blue{Performance with 20 and 5 seeds under 5k SFT on Llama 3.1-8B-Instruct. Metrics follow Table~\ref{tab:main_exp}; full baseline comparisons appear in Appendix~\ref{sec:appendix:limited_seed}.}}
\label{tab:seed_budget_main}
\end{table}

\blue{\textbf{Seed Budget.} Five seeds perform strongly across four task groups under matched settings (Table~\ref{tab:seed_budget_main}). The results remain close to the 20-seed setting on Alpaca, ARC-E, GSM8K, and Text-to-SQL, with larger decreases on ARC-C and MATH. These results support compact exemplar specifications for guiding data preparation. The ARC-Challenge case in Appendix~\ref{sec:appendix:case_study} illustrates this role: the seed pattern guides canonical option and answer representations, while trial-output feedback refines the explanation. Exemplar-derived requirements thus connect structural repair with output refinement. Baseline comparisons, seed resampling, and seed-source transfer results remain in Appendix~\ref{sec:appendix:seed_robustness}.}

\section{Conclusion}
\label{sec:conclusion}

\blue{We propose DataEvolver, an automatic data preparation system based on multi-level self-evolving. It combines seed-guided understanding, operator adaptation, and trial-output feedback to construct and refine reusable data preparation pipelines. Experiments on seven benchmarks demonstrate improvements in downstream LLM performance and preparation efficiency, while the analyses reveal the complementary roles of the two evolving levels. Together, these results establish multi-level self-evolving as an effective approach to automatic data preparation for LLM training.}

% This paper presents DataEvolver, a self-evolving data preparation system for LLM training under seed guidance. DataEvolver combines seed-guided understanding with operator-level and pipeline-level self-evolving to address two practical challenges in automated data preparation: ensuring that generated pipelines are executable and that the resulting data aligns with seed-implied quality requirements. Experiments across diverse benchmarks show that DataEvolver-prepared data yields more reliable improvements than using raw training data or a strong system baseline under the same SFT training scale, and LLM-based judgment further confirms better training-readiness, seed alignment, and explanation quality. 

% In future work, we plan to strengthen robustness across broader data modalities and operator libraries, and explore more principled feedback signals to further improve alignment and stability in self-evolving data preparation.

% Begin the non-main-text limitations section on a fresh page.
\clearpage

% \section*{Acknowledgements}
% We thank all the anonymous reviewers for their insightful and valuable comments. 
% This work was partially supported by the Scientific Research Innovation Capability Support Project for Young Faculty (Grant No. SRICSPYF-ZY2025001) and the National Natural Science Foundation of China (Grant Nos. 62436010, 62441230).
% This work was partially supported by the National Natural Science Foundation of China (Grant Nos. 62436010, 62441230) and the Scientific Research Innovation Capability Support Project for Young Faculty (Grant No. SRICSPYF-ZY2025001).

\section*{Limitations}

In this paper, we propose a self-evolving data preparation system, DataEvolver, which automatically transforms raw data into high-quality data. DataEvolver currently focuses on text-based data preparation for LLMs. Its extensible operator library offers a path toward other modalities, which we leave to future work. Seed examples and LLM-based feedback influence the inferred quality requirements and the evolution process; broader training settings and more reliable task-specific validators warrant further investigation. Comparisons with recent agentic systems focus on mechanisms, while matched empirical evaluations remain future work.

% \input{sections/8_acknowledgements}

\bibliography{references/dataevolver,references/recent2026}

\onecolumn
\appendix
% Keep supplementary paragraphs together without stretched vertical gaps.
\raggedbottom
\section{Interpretation and Related Mechanisms}
\label{sec:appendix:recent_comparison}

\subsection{What Evolves in DataEvolver}
\blue{DataEvolver treats the preparation pipeline as an evolving, reusable program. The evolving artifacts include the operator library, the logical plan, the induced data profile, and the instantiated pipeline. During pipeline construction, the LLM supplies reasoning, generation, and assessment; downstream SFT is performed after the prepared corpus has been produced. Thus, the two evolving levels refer to feedback-driven changes to data preparation artifacts.}

\blue{Seed examples provide a compact specification of the desired supervision, including its structure, explanation style, and quality criteria. The structured profile translates this specification into requirements for pipeline construction. Operator-level feedback addresses missing capabilities and inconsistent dependencies before instantiation. Pipeline-level feedback addresses discrepancies that emerge when the composed program processes actual data. Experience from these trial outputs returns to both data understanding and orchestration, allowing the system to revise its interpretation of the target as well as its transformation strategy. The two loops therefore connect structural feasibility with the quality of the prepared outputs.}

\subsection{Comparison with Recent Agentic Data Systems}
\blue{Recent systems advance automatic data preparation through several complementary mechanisms. DataFlow-Agent synthesizes operators, assembles pipelines, and performs iterative execution verification~\cite{liang2025dataflowllmdrivenframeworkunified}; DataFlow-Harness further grounds code-agent construction in a live operator registry with typed interfaces and schema validation~\cite{he2026dataflowharness}. These capabilities establish that dynamic operator synthesis and execution repair are useful building blocks for agentic data preparation. DataEvolver couples these building blocks with a target specification induced from high-quality examples and with trial-output feedback across two levels of evolution.}

\blue{DataFoundry adapts component policies and prompts after pilot assessment while preserving stable interfaces~\cite{yang2026datafoundry}. DataEvolver couples operator-library expansion and DAG repair with trial-output feedback that revises the seed-induced target profile and subsequent orchestration. CurateEvo evolves curation strategies using downstream development failures after agent post-training~\cite{wang2026curateevo}. Autodata adapts data synthesis strategies using solver feedback~\cite{kulikov2026autodata}. These works highlight the value of assessment signals for evolving data production. In DataEvolver, feedback from sampled pipeline outputs is interpreted relative to the seed-derived target and is accumulated as experience for the next round of understanding and orchestration before full-corpus execution.}

\blue{Recursive harness self-improvement explores co-evolution of task-generation harnesses and solver skills to construct harder valid reasoning tasks~\cite{zhang2026recursiveharness}. DataEvolver addresses the transformation of a given raw corpus into training-ready supervision through an evolving operator library and preparation pipeline. These settings emphasize different evolving artifacts and feedback sources. Together, they motivate continued investigation of self-evolution as a general principle for data systems, with DataEvolver providing a concrete multi-level framework for automatic LLM data preparation.}

\subsection{\blue{Additional Logical Plan Assessment}}
\label{sec:appendix:plan_assessment}
\blue{This existing LLM-rated assessment complements the downstream ablation by measuring operator adequacy, interface consistency, and plan coherence. The main text reports the observed improvements; the full score plot is retained here.}

\begin{figure}[t]
  \centering
  \includegraphics[width=0.55\textwidth]{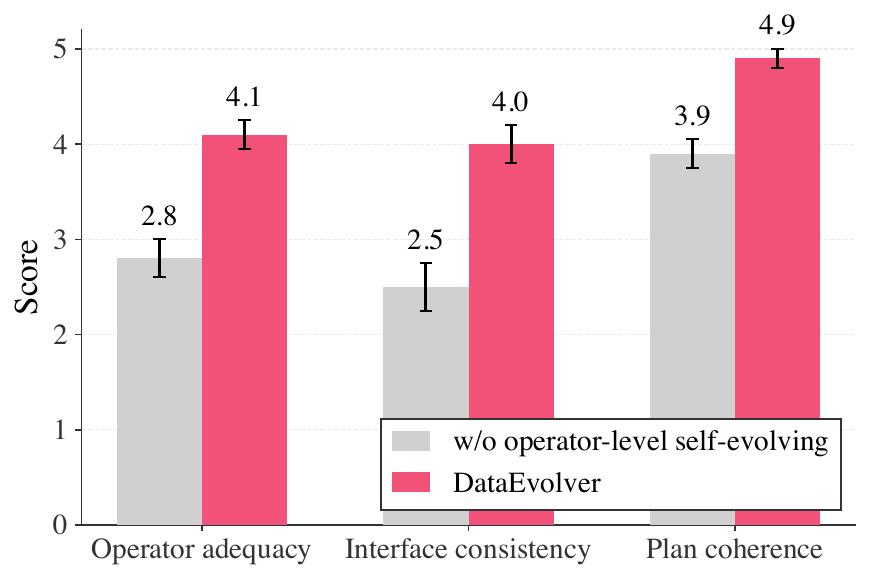}
  \caption{\blue{Operator-level self-evolving improves logical plan quality, yielding higher scores on operator adequacy, interface consistency, and plan coherence.}}
  \label{fig:op_level_analysis}
\end{figure}
\section{Dataset}
\subsection{Raw Data}
\label{sec:appendix:raw_data}

We use seven public benchmarks spanning four task groups: general instruction following (Alpaca), multiple-choice QA (ARC-Easy, ARC-Challenge), math reasoning (GSM8K, MATH), and Text-to-SQL (Spider, BIRD). All raw data are downloaded from HuggingFace using the dataset IDs in Table~\ref{tab:raw_datasets}. We start from the official training split; for GSM8K, we use the main subset (not socratic). For the math group, we follow our adopted setting and mix the two math sources to form a shared raw pool for data preparation, while still reporting GSM8K and MATH separately at evaluation time.

\begin{table*}[t]
\centering
\small
\setlength{\tabcolsep}{3.5pt}
\renewcommand{\arraystretch}{1.08}
\begin{tabular}{@{}l l r@{}}
\toprule
\textbf{Benchmark} & \textbf{HuggingFace ID} & \textbf{\#Train} \\
\midrule
Alpaca & \hflink{tatsu-lab/alpaca} & 52k \\
ARC-Easy & \hflink{allenai/ai2\_arc} & 5.2k \\
ARC-Challenge & \hflink{allenai/ai2\_arc} & 2.59k \\
GSM8K (main) & \hflink{openai/gsm8k} & 8.79k \\
MATH & \hflink{EleutherAI/hendrycks\_math} & 12k \\
Spider & \hflink{xlangai/spider} & 8k \\
BIRD & \hflink{xu3kev/BIRD-SQL-data-train} & 9.43k \\
\bottomrule
\end{tabular}
\caption{Raw training corpora used in this work (HuggingFace snapshots). \#Train denotes the size of the official training split (or subset) used as the raw source before sampling to our SFT budgets.}
\label{tab:raw_datasets}
\end{table*}

\paragraph{Sampling for comparison.}
Our main experiments use two SFT budgets, 1k and 5k, referring to the size of the \emph{final} training set after preparation. To keep the training budget comparable across methods, for datasets with more than 5k training instances, we randomly sample 5k raw instances (with a fixed seed) as the source pool; for datasets below 5k, we use all available training instances as the raw source. All methods (Vanilla SFT, DataFlow, and DataEvolver) draw from the same raw source pool under the same budget.

\paragraph{Why raw data are not training-ready.}
Although these benchmarks provide reliable evaluation signals, their raw training splits are not always directly compatible with a unified, training-ready SFT format. Typical issues include:
\begin{itemize}[leftmargin=1.2em,itemsep=2pt,topsep=2pt]
  \item \textbf{Heterogeneous formats:} instances differ in structure and supervision style across tasks and datasets (e.g., single-turn instruction, multiple-choice options, free-form rationales, or program-like outputs), making it hard to apply a single training template without extra normalization.
  \item \textbf{Underspecified supervision:} many samples omit intermediate structure or constraints implied by the target format, such as explicit reasoning steps, consistent answer normalization, or strict field requirements, which can lead to ambiguous training signals after templating.
  \item \textbf{Noise and redundancy:} a subset of samples contains redundant text, inconsistent structuring, or task-irrelevant artifacts (e.g., boilerplate prompts, duplicated statements, or incomplete fields), which increases variance and can dilute supervision.
  \item \textbf{Inconsistent quality patterns:} even within the same dataset, samples can vary in clarity, completeness, and strictness of formatting, which makes it difficult to obtain stable gains from small-scale SFT subsets.
  \item \textbf{Strong task constraints:} some tasks impose rigid correctness criteria that the raw supervision does not explicitly enforce. For example, text-to-SQL requires executable and schema-consistent outputs, while math reasoning often benefits from structured step decomposition and normalized final answers.
\end{itemize}
These properties motivate seed-driven structured understanding and the two self-evolving loops, which systematically transform raw data into supervision that is both more training-ready and more consistent with the seed-implied target specification.

\subsection{Seed Data}
\label{sec:appendix:seed_data}

% DataEvolver is guided by a small set of manually constructed seed examples (count is \emph{20}). We curated the seeds as lightweight specifications following recent best practices for high-quality SFT data (clear intent, well-formed outputs, and task-faithful formatting), without tuning to test questions or using evaluation signals. Each seed exemplifies a distinct, commonly occurring pattern or a critical reasoning step within its task, thereby providing a foundational template for the system to generalize from. And the seeds make key requirements explicit (e.g., required fields, output structure, and answer normalization) so that DataEvolver can extract a structured target specification and use it to guide pipeline planning and iterative refinement. For tasks where format errors are especially costly (e.g., math reasoning and text-to-SQL), we prioritize seeds that demonstrate canonical reasoning or schema-consistent outputs. Below we show representative examples.

DataEvolver is guided by a small set of manually constructed seed examples (\blue{20 examples in our main setting}). These seeds were curated as lightweight specifications, following recent best practices for high-quality SFT data: clear intent, well-formed outputs, and task-faithful formatting. \blue{Together, these examples encode desired supervision patterns and quality requirements for each target task.} Each seed illustrates a distinct, commonly occurring pattern or a critical reasoning step within its task, providing a foundational template from which the system can generalize. The seeds also include key requirements, such as required fields, output structure, and answer normalization, allowing DataEvolver to extract a structured target specification to guide pipeline planning and iterative refinement. For tasks where format errors are particularly costly (e.g., math reasoning or text-to-SQL), we prioritize seeds that demonstrate canonical reasoning or schema-consistent outputs. Representative examples are shown below.

\begin{seedbox}{}
\textbf{Seed Data - General}
\begin{seedcode}
  "instruction": "Explain why the Earth's temperature fluctuates.",
  "input": "",
  "prompt": "### Instruction:\nExplain why the Earth's temperature fluctuates.\n\n### Response:\n",
  "output": "Earth's temperature fluctuates because the planet's energy balance (incoming sunlight vs. outgoing heat) changes over time.\n\nKey drivers include: 
  (1) seasonal geometry from Earth's axial tilt and orbit, which changes sunlight angle and day length; 
  (2) variations in greenhouse gases and water vapor that affect how much heat is trapped; 
  (3) clouds, aerosols, and volcanic eruptions that reflect sunlight and temporarily cool the surface; 
  (4) ocean-atmosphere circulation (e.g., El Nino/La Nina) that redistributes heat; and 
  (5) surface albedo changes (ice/snow, vegetation, land use) that alter reflectivity.
  
  These factors operate on different time scales (days to decades and beyond), so temperature naturally varies even without a single dominating cause."
\end{seedcode}
\end{seedbox}

\begin{seedbox}{}
\textbf{Seed Data - QA-MCQ}
\begin{seedcode}
  "instruction": "Read the multiple-choice question and select the correct option. Output the option letter and a brief reason.",
  "input": "Question: A boat is acted on by a river current flowing north and by wind blowing on its sails. The boat travels northeast. In which direction is the wind most likely applying force to the sails of the boat?
  Options:
    A. west
    B. east
    C. north
    D. south",
  "output": 
    "Answer: B (east). 
    Reason: The river current provides the northward component, so the wind must supply an eastward force to produce a net motion toward the northeast."
\end{seedcode}
\end{seedbox}
    
\hfill
\hfill

\begin{seedbox}{}
\textbf{Seed Data - Math Reasoning}
\begin{seedcode}
  "instruction": "Solve the math problem. Return (1) an ordered list of steps, one operation per step, and (2) the final answer in LaTeX.",
  "input": "Find all values of x with 0 \\le x < 2\\pi that satisfy \\sin x + \\cos x = \\sqrt{2}. Enter all the solutions, separated by commas.",
  "output": {
    "steps": [
      "Start from sin x + cos x = \\sqrt{2} and square both sides.",
      "Use sin^2 x + cos^2 x = 1 to simplify: 1 + 2 sin x cos x = 2, so sin x cos x = 1/2.",
      "Rewrite 2 sin x cos x as sin 2x, obtaining sin 2x = 1.",
      "Solve on 0 \\le x < 2\\pi: 2x = \\pi/2 or 5\\pi/2, hence x = \\pi/4 or 5\\pi/4.",
      "Check candidates in the original equation; only x = \\pi/4 makes sin x + cos x = \\sqrt{2}."
    ],
    "final_answer": "\\frac{\\pi}{4}"
  }
\end{seedcode}
\end{seedbox}

\begin{seedbox}{}
\textbf{Seed Data - Text-to-SQL}
\begin{seedcode}
  "question": "Among the schools with an SAT excellence rate of over 0.3, what's the highest eligible free rate for students aged 5-17?",
  "db_id": "california_schools",
  "schema_context": {
    "tables": {
      "frpm": ["CDSCode", "Enrollment (Ages 5-17)", "Free Meal Count (Ages 5-17)"],
      "satscores": ["cds", "NumGE1500", "NumTstTakr"]
    },
    "join_keys": [
      {"left": "frpm.CDSCode", "right": "satscores.cds"}
    ]
  },
  "evidence": "Excellence rate = NumGE1500 / NumTstTakr; Eligible free rate (ages 5-17) = Free Meal Count (Ages 5-17) / Enrollment (Ages 5-17).",
  "sql": "SELECT MAX(CAST(t1.`Free Meal Count (Ages 5-17)` AS REAL) / t1.`Enrollment (Ages 5-17)`)\nFROM frpm AS t1\nJOIN satscores AS t2 ON t1.CDSCode = t2.cds\nWHERE CAST(t2.NumGE1500 AS REAL) / t2.NumTstTakr > 0.3;",
  "difficulty": "moderate"
\end{seedcode}
\end{seedbox}

\section{Experimental Setup Details}
\label{sec:appendix:exp_setup}

This appendix summarizes the key experimental configurations for reproducibility, including the training/evaluation toolchains, core SFT hyperparameters, and fixed evaluation settings.

\subsection{Training Configuration}
\label{sec:appendix:training_config}

We run supervised fine-tuning (SFT) with MS-Swift\footnote{\url{https://github.com/modelscope/swift}} and conduct all comparisons under the same backbone-specific training configuration.
Experiments are performed on three backbones: \texttt{Qwen3-8B-Base}, \texttt{Gemma-3-4B-it}, and \texttt{Llama3.1-8B-Instruct}.
Within each backbone, all variants differ only in the constructed training data; all reported results use full-parameter SFT.

We report results at two training scales, 1k and 5k, which denote the number of \emph{final} SFT instances after data preparation. For each backbone and benchmark, compared variants are matched at the same final size. Table~\ref{tab:train_hparams} lists the core hyperparameters shared across benchmarks and variants.

\begin{table}[t]
\centering
\footnotesize
\setlength{\tabcolsep}{5.0pt}
\renewcommand{\arraystretch}{1.08}
\begin{tabular}{@{}lr@{}}
\toprule
Hyperparameter & Value \\
\midrule
Learning rate & $2 \times 10^{-5}$ \\
Per-device batch size & 1 \\
Gradient accumulation steps & 16 \\
Training epochs & 2 \\
Max sequence length & 15{,}000 \\
Warmup ratio & 0.05 \\
Optimizer & AdamW \\
Precision & bfloat16 \\
\bottomrule
\end{tabular}
\caption{Core hyperparameters for full-parameter SFT used in our experiments.}
\label{tab:train_hparams}
\end{table}

\subsection{Evaluation Configuration}
\label{sec:appendix:eval_config}

We evaluate all models with OpenCompass\footnote{\url{https://github.com/open-compass/opencompass}} following the official protocols when available.
Within each benchmark, inference settings are kept fixed across compared variants (e.g., decoding configuration and prompting templates), so that differences can be attributed to the training data.

Table~\ref{tab:infer_settings} summarizes the main inference settings used in evaluation, and Table~\ref{tab:eval_metrics} lists the reported metrics.

\begin{table*}[t]
\centering
\begin{minipage}[t]{0.34\linewidth}
  \centering
  \footnotesize
  \setlength{\tabcolsep}{4.6pt}
  \renewcommand{\arraystretch}{1.08}
  \begin{tabular}{@{}lr@{}}
    \toprule
    Inference setting & Value \\
    \midrule
    Batch size & 32 \\
    Max output length & 2{,}048 \\
    Max sequence length & 2{,}048 \\
    Inference backend & vLLM \\
    GPU memory utilization & 0.6 \\
    Temperature & 0 \\
    \bottomrule
  \end{tabular}
  \caption{Fixed inference settings used by OpenCompass in evaluation.}
  \label{tab:infer_settings}
\end{minipage}%
\hspace{0.02\linewidth}%
\begin{minipage}[t]{0.62\linewidth}
  \centering
  \footnotesize
  \setlength{\tabcolsep}{3.8pt}
  \renewcommand{\arraystretch}{1.08}
  \begin{tabular}{@{}lll@{}}
    \toprule
    Task category & Datasets & Metric \\
    \midrule
    General instruction & Alpaca & AlpacaEval~2.0 win-rate \\
    QA-MCQ & ARC-Easy / ARC-Challenge & Accuracy rate of question \\
    Math reasoning & GSM8K / MATH & Exact match (EM) \\
    Text-to-SQL & Spider / BIRD & Execution accuracy (EX) \\
    \bottomrule
  \end{tabular}
  \caption{Downstream evaluation metrics reported in our experiments.}
  \label{tab:eval_metrics}
\end{minipage}
\end{table*}

\section{Pipeline Statistics}
\label{sec:appendix:pipeline_stats}
This section provides the full pipeline-level statistics used in our efficiency analysis. For each dataset, Table \ref{tab:pipeline_stats} reports (i) \emph{\#Operators}, the total number of operators in the instantiated preparation pipeline; (ii) \emph{\#LLM Operators}, the number of operators that invoke an LLM; and (iii) \emph{Tok/item}, the amortized token consumption per final training instance during the preparation stage. Overall, DataEvolver reduces token cost per item across all datasets (40.3\% macro-average), even when it uses more fine-grained non-LLM operator steps for structured transformation. The savings are most visible on instruction and MCQ tasks (e.g., Alpaca: 3,214$\rightarrow$1,218; ARC-Easy: 3,768$\rightarrow$1,647), while Text-to-SQL shows smaller but consistent reductions due to stronger structural constraints.

\begin{table*}[t]
\centering
\footnotesize
\setlength{\tabcolsep}{3.2pt}
\renewcommand{\arraystretch}{1.08}
\begin{tabular}{lccc ccc c}
\toprule
& \multicolumn{3}{c}{\textbf{DataFlow}} & \multicolumn{3}{c}{\textbf{DataEvolver}} & \textbf{$\Delta$ Tok/item $\downarrow$} \\
\cmidrule(lr){2-4}\cmidrule(lr){5-7}\cmidrule(lr){8-8}
\textbf{Dataset} & \textbf{\#Operators} & \textbf{\#LLM Operators} & \textbf{Tok/item}
& \textbf{\#Operators} & \textbf{\#LLM Operators} & \textbf{Tok/item}
& \textbf{(\%)} \\
\midrule
Alpaca & 3 & 3 & 3,214 & 9 & 4 & 1,218 & 62.1 \\
ARC-Easy & 5 & 4 & 3,768 & 7 & 3 & 1,647 & 56.3 \\
ARC-Challenge & 5 & 4 & 4,091 & 9 & 4 & 2,130 & 47.9 \\
GSM8K & 10 & 6 & 2,564 & 7 & 3 & 1,864 & 27.3 \\
MATH & 10 & 6 & 4,832 & 10 & 5 & 2,958 & 38.8 \\
Spider & 7 & 4 & 2,592 & 8 & 4 & 1,919 & 26.0 \\
BIRD & 7 & 4 & 2,755 & 8 & 4 & 2,111 & 23.4 \\
\midrule
Macro avg & 6.7 & 4.4 & 3,402 & 8.3 & 3.9 & 1,978 & 40.3 \\
\bottomrule
\end{tabular}
\caption{Pipeline statistics for data preparation. \emph{\#Operators} counts the total number of the operators in the instantiated pipeline. \emph{\#LLM Operators} counts the number of the operators that invoke an LLM (i.e., call LLM API within the operator). \emph{Tok/item} indicates the amortized token consumption per final training instance during preparation. $\Delta$ shows the relative reduction of DataEvolver compared to DataFlow, macro-averaged over datasets.}
\label{tab:pipeline_stats}
\end{table*}

\subsection{Generation Details}
\label{sec:appendix:op_details}

Tables~\ref{tab:predefined_operators} and~\ref{tab:op_generation_details} list the predefined and generated operators. We start from a fixed set of 21 predefined, task-agnostic operators (I/O, field editing, normalization/cleaning, and a few LLM-assisted primitives). While this inventory supports generic transformations, it often lacks task-aware capabilities needed to convert raw corpora into the seed-aligned supervision format.

Operator-level self-evolving fills these capability gaps by synthesizing a small number of missing operators conditioned on the seed-driven specification. The generated operators act as bridges that enforce critical constraints that would otherwise be brittle to express via long compositions. As shown in the table, the synthesized operators are task-specific: for General instruction and QA-MCQ, they focus on constructing standardized instances and enforcing answer/format consistency; for Math reasoning, they add normalization, step structuring, and final-answer verification; for Text-to-SQL, they emphasize schema grounding (e.g., building/pruning schema context and fusing evidence). Overall, this table illustrates that DataEvolver can minimally extend a shared operator inventory to better match task requirements, reducing manual per-task engineering.

\begin{table*}[t]
\centering
\footnotesize
\setlength{\tabcolsep}{3.5pt}
\begin{tabular}{@{}p{0.72\textwidth}p{0.22\textwidth}@{}}
\toprule
\textbf{Predefined operator signature} & \textbf{Output} \\
\midrule
read\_data(file\_path) & data\_list \\
write\_data(data\_list,file\_path) & data\_list \\
extract\_field(data,field\_name) & value \\
add\_field(data,field\_name,field\_value) & data \\
remove\_field(data,field\_name) & data \\
merge\_fields(data,source\_fields,target\_field) & data \\
split\_field(data,source\_field,target\_fields) & data \\
transform\_field(data,field\_name,rule) & data \\
filter\_data(data,condition) & data \\
deduplicate(data,key\_fields) & data \\
sample\_data(data,sample\_size,method) & data \\
normalize\_format(data,rules) & data \\
clean\_data(data,rules) & data \\
validate\_format(data,format\_spec) & (report,data) \\
combine\_sources(data1,data2,strategy) & data \\
sort\_data(data,sort\_key,sort\_order) & data \\
aggregate\_data(data,aggregation\_field,aggregation\_type) & data \\
format\_output(data,target\_schema) & data \\
call\_llm\_for\_generation(data,input\_field,instruction) & data \\
call\_llm\_for\_extraction(data,source\_field,target\_structure) & data \\
call\_llm\_for\_evaluation(data,content\_field,criteria) & (score,data) \\
\bottomrule
\end{tabular}
\caption{Predefined operator set shared across tasks.}
\label{tab:predefined_operators}
\end{table*}

\begin{table*}[t]
\centering
\footnotesize
\setlength{\tabcolsep}{3.5pt}
\begin{tabular}{@{}p{0.18\textwidth}p{0.76\textwidth}@{}}
\toprule
\textbf{Task} & \textbf{New operator signatures and outputs} \\
\midrule
General & BuildInstPrompt(instruction,input,style\_spec) $\rightarrow$ prompt\newline StripBoilerplate(text,patterns) $\rightarrow$ (instruction,input,output) \\
\midrule
QA–MCQ & BuildMCQInstance(question,choices\{label,text\}) $\rightarrow$ options\newline AlignGoldToChoice(answerKey,options) $\rightarrow$ (answer,answer\_text) \\
\midrule
Math Reasoning & NormalizeMath(tex\_str,mode) $\rightarrow$ norm\_tex\newline SplitIntoSteps(solution\_tex,granularity) $\rightarrow$ steps\newline VerifyFinalAnswer(problem,steps,answer\_field) $\rightarrow$ (final\_answer,consistency\_check) \\
\midrule
Text-to-SQL & SchemaContext(db\_id,schema\_cache) $\rightarrow$ schema\_ctx\newline PruneSchema(question,schema\_ctx,top\_k) $\rightarrow$ pruned\_schema\newline FuseEvidence(question,evidence,pruned\_schema) $\rightarrow$ prompt\_ctx \\
\bottomrule
\end{tabular}
\caption{Task-specific operators newly generated during self-evolving. The predefined operator set shared across tasks is listed in Table~\ref{tab:predefined_operators}.}
\label{tab:op_generation_details}
\end{table*}

\section{Case Study}
\label{sec:appendix:case_study}

\begin{figure*}[t]
    \centering
    \includegraphics[width=1\linewidth]{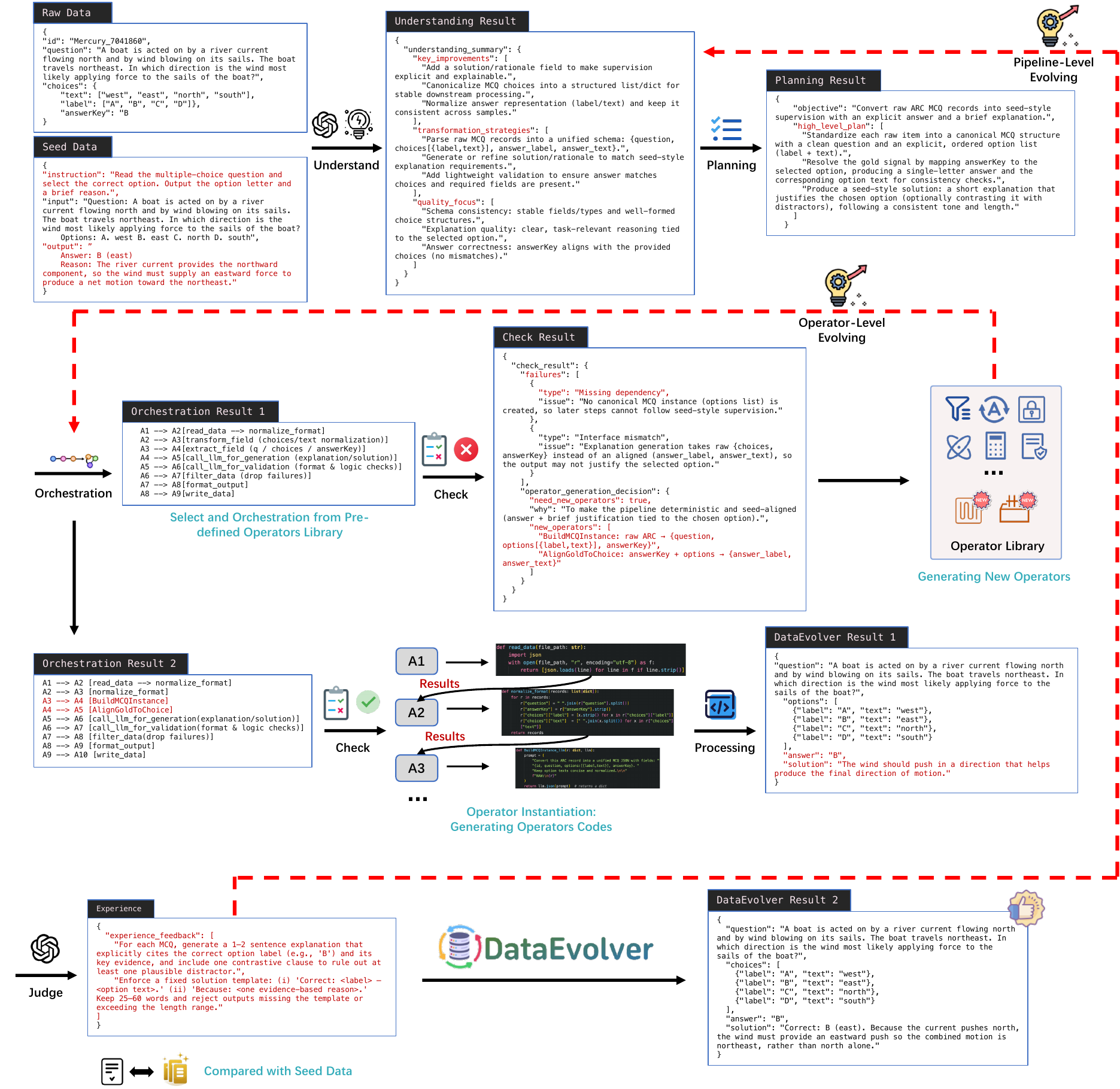}
    \caption{ARC-Challenge case study. DataEvolver induces a pipeline from seed-style supervision and diagnoses capability gaps in an initial plan built with predefined operators. It then synthesizes two operators to re-orchestrate a seed-aligned pipeline, and uses judge-driven experience feedback to refine the output style. For readability, we run the induced pipeline on the same raw example shown in the figure.}
    \label{fig:case_1}
\end{figure*}

\subsection{End-to-End Example (ARC-Challenge)}
Figure~\ref{fig:case_1} shows a minimal end-to-end walkthrough on \textsc{ARC-Challenge} (QA--MCQ). Starting from one raw ARC record and a seed-style supervision target (an explicit option label plus a brief explanation), DataEvolver first summarizes the raw-to-seed gap via structured understanding (e.g., missing supervision fields and format constraints) and then produces a high-level plan that specifies what to generate and how the final output should look.

Based on this plan, DataEvolver orchestrates a logical pipeline from the predefined operator inventory. The \texttt{Check} stage then diagnoses capability gaps that prevent the pipeline from reliably matching the seed-style interface---in particular, the raw record needs to be converted into a canonical MCQ instance, and the gold \texttt{answerKey} must be aligned to a stable answer representation. This triggers operator-level self-evolving to synthesize \texttt{BuildMCQInstance} (canonical MCQ instance construction) and \texttt{AlignGoldToChoice} (aligning \texttt{answerKey} to a consistent answer interface). With the expanded operator set, DataEvolver re-orchestrates the pipeline and incorporates experience feedback to enforce a seed-like explanation pattern, yielding prepared data that is closely aligned with the seed supervision style.

\subsection{DAG and Instantiation Example (MATH)}

\begin{figure*}[t]
    \centering
    \includegraphics[width=\linewidth]{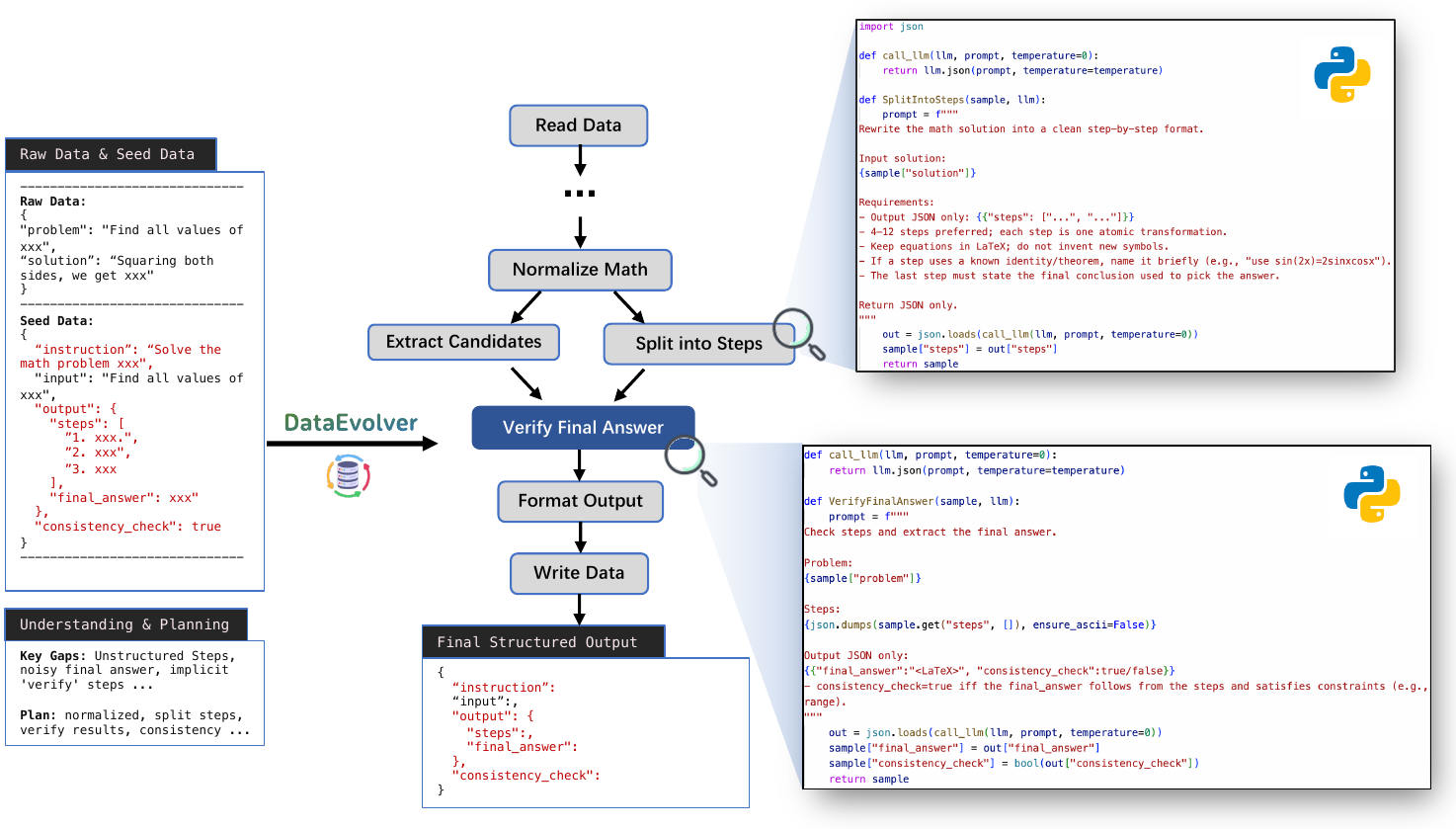}
    \caption{\textsc{MATH} case study highlighting (i) a DAG-style orchestration with branching/converging steps and (ii) operator instantiation. Starting from raw \{\texttt{problem}, \texttt{solution}\}, DataEvolver targets seed-style supervision with structured \texttt{steps} and a verified \texttt{final\_answer}. The pipeline introduces task-specific operators (e.g., \texttt{SplitIntoSteps}, \texttt{VerifyFinalAnswer}) and instantiates them as LLM-calling code to produce a consistency-checked output.}
    \label{fig:case_2}
\end{figure*}

Figure~\ref{fig:case_2} presents a \textsc{MATH} example that emphasizes orchestration structure and operator instantiation. Unlike the linear walkthrough in Figure~\ref{fig:case_1}, this pipeline forms a lightweight DAG: after normalizing the input, it splits into two branches for (i) extracting answer candidates and (ii) restructuring the solution into step-wise supervision, which later merge for final verification.

This case highlights operator-level self-evolving and instantiation in one view. DataEvolver synthesizes math-specific operators to meet seed-style requirements: \texttt{SplitIntoSteps} produces clean, atomic steps, while \texttt{VerifyFinalAnswer} verifies the extracted \texttt{final\_answer} against the steps and outputs a boolean \texttt{consistency\_check}. The right panel shows the instantiated LLM-calling code snippets, illustrating how logical operators are concretized into executable components for data transformation.

\subsection{Raw-to-Prepared Data Examples}

\begin{figure*}[t]
    \centering
    \includegraphics[width=1\linewidth]{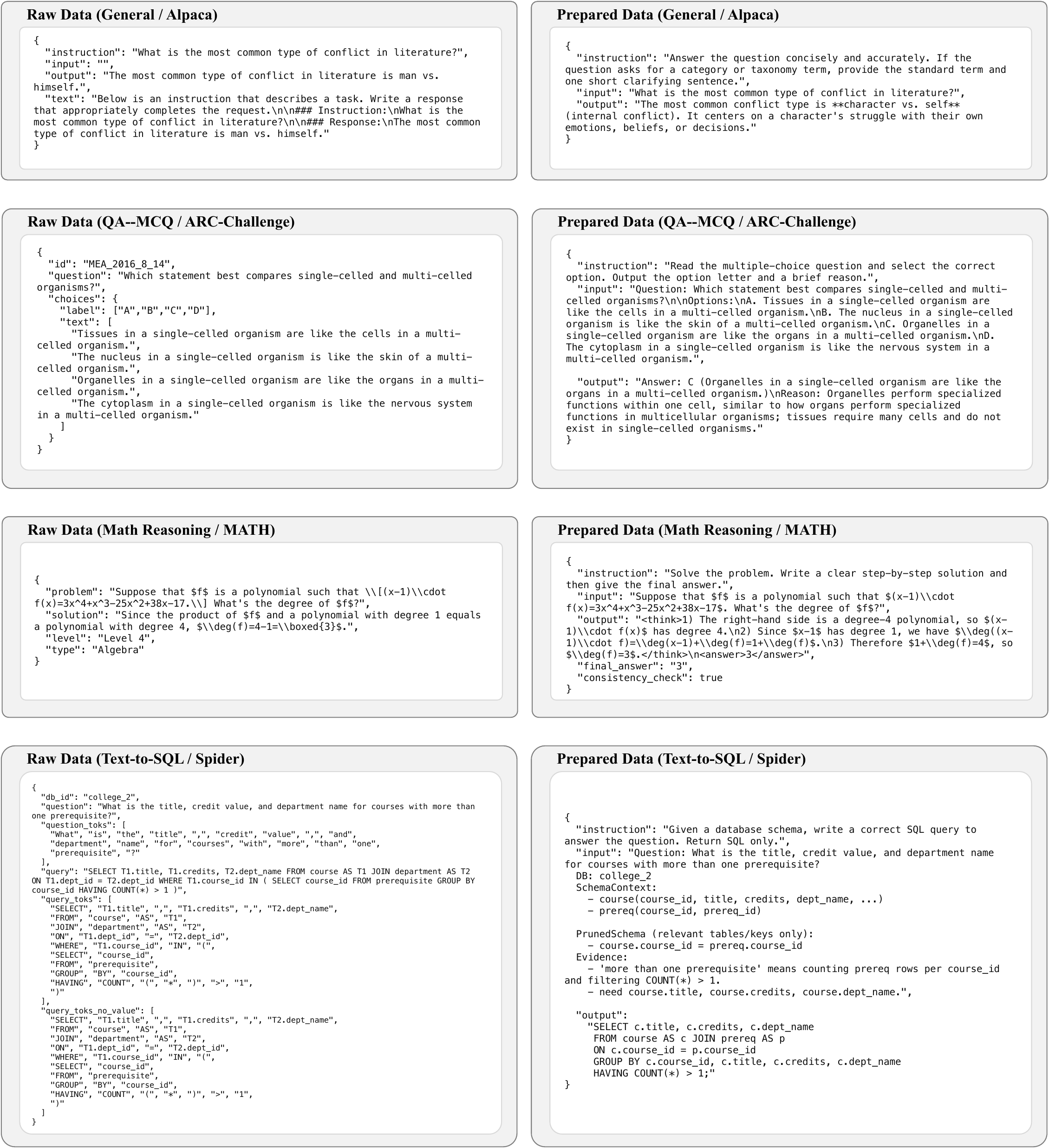}
    \caption{Raw-to-prepared data examples across four task categories in our paper: General, QA–MCQ, Math Reasoning and Text-to-SQL.}
    \label{fig:case_3_1234}
\end{figure*}

To make DataEvolver’s data transformation more concrete, we provide representative raw-to-prepared examples across task categories (Figure~\ref{fig:case_3_1234}). Each example shows one raw record and its corresponding prepared training instance produced by the final pipeline, highlighting canonicalized interfaces, seed-aligned supervision fields, and consistent training formats.

\clearpage

% sec E

\section{Robustness to Seed Specification}
\label{sec:appendix:seed_robustness}

This section supplements the main paper with additional analyses on the role of seed data in DataEvolver. Since seeds serve as lightweight specifications of target supervision format and quality criteria, an important question is whether DataEvolver is sensitive to the number or exact choice of seed examples. To this end, we provide three additional validations: (i) performance under limited seed supervision, (ii) robustness to different re-sampled seed groups, and (iii) \blue{seed-source transfer across reasoning datasets}.

% This section supplements the main paper with additional analyses on the role of seed data in DataEvolver. Since seeds serve as lightweight specifications of the target data distribution, an important question is whether DataEvolver is sensitive to the number or exact choice of seed examples. To this end, we provide three additional validations: (i) performance under limited seed supervision, (ii) robustness to different re-sampled seed groups, and (iii) cross-seed held-out validation across reasoning datasets.

\subsection{Performance under Limited Seed Supervision}
\label{sec:appendix:limited_seed}

To assess whether DataEvolver remains effective with only a very small seed set, we reduce the number of seed examples from 20 (used in the main experiments) to 5, while keeping the backbone and SFT budget unchanged. Specifically, we use \texttt{Llama3.1-8B-Instruct} with a 5k-item SFT budget and compare the resulting downstream performance against Vanilla-SFT, DataFlow-SFT, and the standard DataEvolver-SFT setting with 20 seeds.

\begin{table*}[t]
\centering
\small
\begin{tabular}{lccccccc}
\toprule
Method 
& Alpaca 
& ARC-E 
& ARC-C 
& GSM8K 
& MATH 
& Spider 
& BIRD \\
& (Win-rate) & (Acc.) & (Acc.) & (EM) & (EM) & (Exec-Acc) & (Exec-Acc) \\
\midrule
Vanilla-SFT (raw)          & 39.4 & 86.19 & 87.21 & 89.43 & 52.33 & 61.92 & 43.18 \\
DataFlow-SFT               & 48.0 & 96.03 & 90.88 & 88.53 & 64.58 & 71.39 & 49.05 \\
DataEvolver-SFT (20 seeds) & 52.5 & 98.47 & 92.03 & 89.34 & 67.74 & 72.66 & 52.85 \\
DataEvolver-SFT (5 seeds)  & 51.4 & 98.02 & 90.24 & 89.02 & 64.98 & 72.45 & 51.73 \\
\bottomrule
\end{tabular}
\caption{Performance of DataEvolver under limited seed supervision (\texttt{Llama3.1-8B-Instruct}, 5k SFT budget). The main setting uses 20 seeds, while the additional row reports results with only 5 seeds under the same backbone and SFT budget.}
\label{tab:limited_seed}
\end{table*}

The results in Table~\ref{tab:limited_seed} show that DataEvolver remains highly effective even with only 5 seeds. Although reducing the seed size from 20 to 5 leads to moderate performance drops on some tasks, the 5-seed setting \blue{improves over Vanilla-SFT on six of the seven benchmarks} and surpasses DataFlow-SFT on most benchmarks. This suggests that DataEvolver does not require a large seed set to function well; instead, a handful of high-quality seeds is already sufficient to provide a strong quality anchor for data preparation. In practice, the 20-seed setting used in the main paper can therefore be viewed as a practical trade-off between annotation cost and robustness.

\subsection{Robustness to Seed Set Resampling}
\label{sec:appendix:seed_resampling}

Beyond the number of seeds, another concern is whether DataEvolver is overly dependent on one particular seed group. To test this, we keep the seed count fixed at 20, re-sample different seed sets, and re-run the full data preparation and SFT pipeline under the same configuration as the main experiment. We again use \texttt{Llama3.1-8B-Instruct} with a 5k-item SFT budget.

\begin{table*}[t]
\centering
\small
\begin{tabular}{lccccccc}
\toprule
Setting (20 seeds) 
& Alpaca 
& ARC-E 
& ARC-C 
& GSM8K 
& MATH 
& Spider 
& BIRD \\
& (Win-rate) & (Acc.) & (Acc.) & (EM) & (EM) & (Exec-Acc) & (Exec-Acc) \\
\midrule
Seed set A               & 52.5 & 98.47 & 92.03 & 89.34 & 67.74 & 72.66 & 52.85 \\
Seed set B               & 51.9 & 98.12 & 91.88 & 89.19 & 66.08 & 72.84 & 53.06 \\
Seed set C               & 52.3 & 98.32 & 92.18 & 89.49 & 67.21 & 71.52 & 52.43 \\
\midrule
\shortstack{Mean} & \shortstack{52.23} & \shortstack{98.30} & \shortstack{92.03} & \shortstack{89.34} & \shortstack{67.01} & \shortstack{72.34} & \shortstack{52.78} \\
\shortstack{$\pm$ Std.} & \shortstack{$\pm$ 0.31} & \shortstack{$\pm$ 0.18} & \shortstack{$\pm$ 0.15} & \shortstack{$\pm$ 0.15} & \shortstack{$\pm$ 0.85} & \shortstack{$\pm$ 0.72} & \shortstack{$\pm$ 0.32} \\
\bottomrule
\end{tabular}
\caption{Robustness to different re-sampled seed groups. All runs use the same backbone, SFT budget, and seed count (20), while varying the specific seed examples. Seed set A is the configuration reported in Table~\ref{tab:main_exp}, and Seed sets B and C are re-sampled from 20 seeds.}
\label{tab:seed_resampling}
\end{table*}

As shown in Table~\ref{tab:seed_resampling}, performance remains stable across different seed groups, with low variance on most datasets. This indicates that DataEvolver is not locked to one specific seed pattern, but instead abstracts more general quality specifications from seeds and applies them consistently during pipeline evolution. These results support our design choice of using seeds as lightweight specifications rather than templates to be copied.

\subsection{\blue{Seed-source Transfer}}
\label{sec:appendix:cross_seed}

\blue{We examine transfer of seed-derived requirements between \textsc{GSM8K} and \textsc{MATH}. Following the math setup, both datasets contribute to a shared reasoning raw pool, while the seed examples come from only one of the two sources. We evaluate downstream SFT performance on both benchmarks using \texttt{Llama3.1-8B-Instruct} and a 5k-item SFT budget.}

% To further test whether DataEvolver over-specializes to the particular phenomena covered by the seeds, we perform a targeted cross-seed held-out validation on two reasoning datasets with noticeably different characteristics: \textsc{GSM8K} and \textsc{MATH}. Concretely, we use the seed data from one dataset to drive DataEvolver, but evaluate the downstream SFT performance on both datasets. All results are obtained with \texttt{Llama3.1-8B-Instruct} and a 5k-item SFT budget.

\begin{table*}[t]
\centering
\small
\begin{tabular}{lcc}
\toprule
Seed Dataset Used for Evolution & GSM8K (EM) & MATH (EM) \\
\midrule
Seed data from \textsc{GSM8K} & 89.34 & 66.80 \\
Seed data from \textsc{MATH}  & 89.15 & 67.74 \\
\bottomrule
\end{tabular}
\caption{\blue{Seed-source transfer on reasoning tasks. The raw pool contains both \textsc{GSM8K} and \textsc{MATH}; seeds come from one source, and evaluation covers both benchmarks.}}
\label{tab:cross_seed}
\end{table*}

\blue{Table~\ref{tab:cross_seed} shows strong performance with either seed source. In this shared-pool setting, requirements induced from one reasoning source remain useful for supervision evaluated on both benchmarks. The results support transfer of seed-derived reasoning structure, format consistency, and answer constraints between these related tasks.}

% sec f
\section{Additional Validation and Practicality Analyses}
\label{sec:appendix:additional_validation}

This section provides additional validations beyond the main paper, focusing on two complementary aspects. First, we validate the reliability of the LLM judge used in both pipeline evolution and data-quality analysis. Second, we supplement the efficiency analysis with deployment-oriented measurements, including end-to-end wall time and a token-matched SFT control to better separate the effects of data quality from token volume.

\subsection{Human Validation of the LLM Judge}
\label{sec:appendix:judge_validation}

To validate the reliability of the LLM judge used in our analysis, we conduct a small-scale human evaluation on sampled prepared items from four task groups. For each task group, we randomly sample 50 prepared instances and ask two annotators to rate their quality on a 1--5 Likert scale, considering correctness, completeness, clarity, and compliance with format or task constraints. We then compare the average human score with the average judge score and report Spearman's $\rho$ between them.

\begin{table*}[t]
\centering
\small
\setlength{\tabcolsep}{4pt}
\begin{tabular}{lccc}
\toprule
Task & \#Items & Human avg. & Judge avg. \\
\midrule
Instruction (Alpaca)      & 50 & 4.62 & 4.70 \\
QA--MCQ (ARC)             & 50 & 4.14 & 4.18 \\
Math (GSM8K/MATH)         & 50 & 4.62 & 4.82 \\
Text-to-SQL (Spider/BIRD) & 50 & 4.39 & 4.45 \\
\bottomrule
\end{tabular}
\caption{Human and judge average scores on sampled prepared items.}
\label{tab:judge_human_avg}
\end{table*}

\begin{table}[t]
\centering
\small
\setlength{\tabcolsep}{5pt}
\begin{tabular}{lc}
\toprule
Task & Human--Judge $\rho$ \\
\midrule
Instruction (Alpaca)      & 0.90 \\
QA--MCQ (ARC)             & 0.88 \\
Math (GSM8K/MATH)         & 0.84 \\
Text-to-SQL (Spider/BIRD) & 0.91 \\
\bottomrule
\end{tabular}
\caption{Human--judge correlation on sampled prepared items, measured by Spearman's $\rho$.}
\label{tab:judge_human_corr}
\end{table}

The results in Tables~\ref{tab:judge_human_avg} and~\ref{tab:judge_human_corr} show strong agreement between human judgments and the LLM judge across all task groups. This supports our use of the judge as an efficient feedback signal during self-evolving and as a scalable proxy for data-quality analysis. Importantly, the core effectiveness claims in the main paper still rely on objective downstream metrics such as execution accuracy and exact match, rather than on judge scores alone.

\subsection{Wall-time Efficiency}
\label{sec:appendix:wall_time}

While token consumption provides a hardware-agnostic efficiency metric, it does not directly reflect the latency of a complete data-preparation run in practice. To supplement the token-cost analysis in the main paper, we additionally measure the end-to-end wall time of the data preparation stage under a controlled deployment setting. Specifically, we use API-based inference with concurrency $=1$ and the same generation model for both methods. Both methods operate on the same raw corpus and produce the same final prepared size (5k items), each under its default preparation procedure.

% While token consumption provides a hardware-agnostic efficiency metric, it does not directly reflect the latency of a complete data-preparation run in practice. To supplement the token-cost analysis in the main paper, we additionally measure the end-to-end wall time of the data preparation stage under a controlled deployment setting. Specifically, we use API-based model inference with concurrency $=1$, the same generation backbone as in the main experiments, the same raw corpus, the same final prepared size (5k items), and the same operator search and pipeline evolution configuration.

\begin{table*}[t]
\centering
\small
\setlength{\tabcolsep}{4pt}
\begin{tabular}{lccc}
\toprule
Method & Tok/item & Total Tokens & Wall Time (h) \\
\midrule
DataFlow    & 3,402 & 17,013,682 & $\approx$ 2.52 \\
DataEvolver & 1,983 & 9,915,406  & $\approx$ 1.63 \\
\bottomrule
\end{tabular}
\caption{End-to-end wall-time comparison under a controlled API-based setting (concurrency $=1$).}
\label{tab:wall_time}
\end{table*}

As shown in Table~\ref{tab:wall_time}, DataEvolver not only reduces token consumption, but also achieves a clear advantage in end-to-end wall time. This indicates that the proposed multi-level self-evolving pipeline is practical not only from a token-efficiency perspective, but also in terms of engineering latency. Since absolute wall time depends on deployment configuration and system environment, we report it here as a complementary deployment-oriented measure rather than as a replacement for token cost.

\subsection{Token-matched SFT Control}
\label{sec:appendix:token_matched}

In the main experiments, training budgets are controlled primarily by the number of SFT instances (e.g., 1k or 5k). However, because prepared data may differ in average length from raw data, equalizing the number of instances does not strictly guarantee identical total training tokens. To better isolate the effect of data quality from token quantity, we perform a representative token-matched SFT control. Concretely, we first compute the total token count of the prepared training set and then subsample the raw training data so that both conditions use the same total training tokens.

\begin{table*}[ht]
\centering
\small
\setlength{\tabcolsep}{4pt}
\begin{tabular}{lccc}
\toprule
Setting & Total Tokens & Alpaca & ARC-E / ARC-C \\
\midrule
Vanilla-SFT (raw, token-matched)  & 1,500,000 & 39.3 & 86.5 / 89.0 \\
DataEvolver-SFT (prepared, token-matched) & 1,500,000 & 52.1 & 98.4 / 92.2 \\
\bottomrule
\end{tabular}
\caption{Representative token-matched SFT results. Both settings use the same total number of training tokens.}
\label{tab:token_matched}
\end{table*}

The results in Table~\ref{tab:token_matched} show that DataEvolver's gains persist even under an equal-token training budget. This suggests that the improvements are not merely due to using more training tokens, but are primarily driven by better prepared training data. We report a representative subset here for brevity. %The results already suggest that the gains are primarily driven by better prepared training data rather than by increased token volume.

% We report this as a representative control here, and leave a larger-scale token-matched comparison across all benchmarks for future work.

\clearpage

\section{Prompt Strategy}
\label{sec:appendix:prompts}

This appendix section reports the key prompt templates used by DataEvolver. We include the core prompts corresponding to the four main functions in our method, which drive structured understanding, planning/orchestration, and pipeline-level self-evolving (quality judging and experience generation). We report only the core prompts used in the main workflow.
% Additional prompt variants and implementation details are provided in the project files.

\begin{tcolorbox}[
  title=Prompt for Structured Understanding,
  colback=blue!8,
  colframe=blue!55!black,
  arc=1mm,
  boxrule=0.9pt,
  left=1mm,right=1mm,top=1mm,bottom=1mm,
  fonttitle=\scriptsize,
  breakable
]
\footnotesize

\textbf{System Prompt.}\\
You are an expert in data analysis and understanding. Your task is to analyze raw data and seed data to produce a comprehensive structured understanding.

Key requirements:
\begin{itemize}\setlength{\itemsep}{1pt}\setlength{\topsep}{2pt}
  \item Output \textbf{ONLY valid JSON}, with no additional text.
  \item Be dataset-agnostic: focus on general patterns, not dataset names.
  \item Provide detailed, actionable insights for downstream pipeline construction.
  \item Pay special attention to \textbf{nested structures}: analyze all fields including nested objects and arrays, and describe their types, meanings, and relationships.
\end{itemize}

\vspace{4pt}
\textbf{User Prompt Template.}\\
Analyze the following raw data and seed data to produce a structured understanding. Pay special attention to nested structures (objects/arrays) and analyze all fields.

\vspace{3pt}
\textbf{Pipeline Configuration}\\
\texttt{\{pipeline\_config\}}

\vspace{3pt}
\textbf{User Requirements / Task Description (Optional)}\\
\texttt{\{user\_requirements\}}

\vspace{3pt}
\textbf{Raw Data Preview}\\
\texttt{\{raw\_data\_preview\}}

\vspace{3pt}
\textbf{Seed Data Preview}\\
\texttt{\{seed\_data\_preview\}}

\vspace{4pt}
\textbf{Output Requirement.}\\
Return \textbf{ONLY valid JSON} with the following structure:

\vspace{2pt}
\noindent
\texttt{\{}\\
\texttt{\ \ "language": "primary language inferred from raw/seed text",}\\
\texttt{\ \ "format\_summary": \{}\\
\texttt{\ \ \ \ "raw\_format": "JSONL/JSON/CSV/TXT/etc",}\\
\texttt{\ \ \ \ "seed\_format": "JSONL/JSON/CSV/TXT/etc",}\\
\texttt{\ \ \ \ "key\_differences": ["format/style differences that matter for training"],}\\
\texttt{\ \ \ \ "nested\_schema\_notes": ["nested objects/arrays and their field paths/types"]}\\
\texttt{\ \ \},}\\
\texttt{\ \ "spec\_from\_seeds": \{}\\
\texttt{\ \ \ \ "required\_fields": ["required output fields implied by seeds"],}\\
\texttt{\ \ \ \ "format\_rules": ["must-follow formatting rules implied by seeds"],}\\
\texttt{\ \ \ \ "quality\_criteria": ["quality expectations implied by seeds"]}\\
\texttt{\ \ \},}\\
\texttt{\ \ "gap\_diagnosis": \{}\\
\texttt{\ \ \ \ "major\_gaps": ["what raw data lacks vs seeds (fields/structure/steps/noise/redundancy)"],}\\
\texttt{\ \ \ \ "common\_failure\_patterns": ["recurring issues observed from raw preview"]}\\
\texttt{\ \ \},}\\
\texttt{\ \ "processing\_targets": ["actionable targets for the pipeline to reach seed quality"],}\\
\texttt{\ \ "transformation\_plan": ["high-level steps to transform raw data towards seed style/spec"]}\\
\texttt{\}}

\end{tcolorbox}

\begin{tcolorbox}[
  title=Prompts for Operator-Level Self-Evolving,
  colback=blue!8,
  colframe=blue!55!black,
  arc=1mm,
  boxrule=0.9pt,
  left=1mm,right=1mm,top=1mm,bottom=1mm,
  fonttitle=\scriptsize,
  breakable
]
\footnotesize

\textbf{Output rule:} return \emph{only} valid JSON.

\vspace{4pt}
\textbf{Stage 3: Pipeline Orchestration ($f_{\text{orch}}$)}

\textbf{System Prompt.}\\
You are an expert in composing executable \emph{logical} data-processing pipelines. Given (1) an optimization blueprint, (2) a structured understanding (specification), and (3) an operator inventory, produce a concise pipeline plan. The plan must start with \texttt{read\_data} and end with \texttt{write\_data}. Ensure data-flow coherence (each input is produced by an upstream output) and avoid redundant steps. Output valid JSON only.

\textbf{User Prompt Template.}\\
\textit{Compose a logical operator pipeline.}

\textbf{Optimization blueprint}\\
\texttt{\{optimization\_blueprint\}}

\textbf{Structured understanding (specification)}\\
\texttt{\{understanding\_result\}}

\textbf{Operator inventory}\\
\texttt{\{operator\_registry\}}

Return JSON:
\begin{lstlisting}[basicstyle=\ttfamily\footnotesize,columns=fullflexible,breaklines=true,breakatwhitespace=false,showstringspaces=false]
{
  "pipeline_dag": {
    "nodes": [
      {
        "step_id": "step_1",
        "operator": "read_data",
        "input_keys": ["file_path"],
        "output_keys": ["data_list"],
        "parameters": {},
        "depends_on": [],
        "description": "..."
      }
    ],
    "edges": [["step_1","step_2"]]
  }
}
\end{lstlisting}

\vspace{6pt}
\textbf{Stage 4: Closure Check ($f_{\text{check}}$)}

\textbf{System Prompt.}\\
You verify whether the pipeline plan is executable at the logical level. Check: (i) interface consistency (required inputs exist and match upstream outputs), (ii) dependency completeness (required intermediate artifacts/fields are produced before use), and (iii) capability coverage (no missing operator capability). If issues exist, propose minimal repairs; if capabilities are missing, request new operators with clear signatures. Output valid JSON only.

\textbf{User Prompt Template.}\\
\textit{Check the pipeline plan and decide whether new operators are needed.}

\textbf{Pipeline plan}\\
\texttt{\{pipeline\_dag\}}

\textbf{Structured understanding (specification)}\\
\texttt{\{understanding\_result\}}

\textbf{Operator inventory}\\
\texttt{\{operator\_registry\}}

Return JSON:
\begin{lstlisting}[basicstyle=\ttfamily\footnotesize,columns=fullflexible,breaklines=true,breakatwhitespace=false,showstringspaces=false]
{
  "is_executable": true,
  "issues": [
    {
      "type": "missing_dependency|interface_mismatch|dag_disconnected|ordering_conflict",
      "where": "step_k",
      "details": "..."
    }
  ],
  "repair_suggestions": [{"action": "...", "details": "..."}],
  "need_new_operators": false,
  "new_operator_specs": [
    {
      "name": "NewOperatorName",
      "description": "...",
      "input_keys": ["..."],
      "output_keys": ["..."],
      "motivation": "..."
    }
  ]
}
\end{lstlisting}

\vspace{4pt}
\textbf{Control flow.} Run orchestration, then closure check. If \texttt{need\_new\_operators=true}, synthesize the requested operators, update the inventory, and re-run orchestration.

\end{tcolorbox}

\begin{tcolorbox}[
  title=Prompts for Pipeline-level Self-Evolving,
  colback=blue!8,
  colframe=blue!55!black,
  arc=1mm,
  boxrule=0.9pt,
  left=1mm,right=1mm,top=1mm,bottom=1mm,
  fonttitle=\scriptsize,
  breakable
]
\footnotesize

\textbf{Output rule:} the quality check returns valid JSON only; the experience mining returns plain text only.

\vspace{4pt}
\textbf{(1) Quality Check Prompt ($f_{\text{qc}}$)}

\textbf{System Prompt.}\\
You are a data quality assessment expert. Compare the \emph{processed} dataset produced by the current pipeline with the \emph{seed} dataset that represents the desired training style and constraints. The two sets are not paired by index; infer \emph{dataset-level} patterns and distributional differences. Focus on four aspects:
(1) implicit requirements shown by seeds but missing in processed data;
(2) style and format consistency (structure, tone, length, required sections);
(3) recurring error patterns in processed data (format violations, missing fields, inconsistencies);
(4) difficulty distribution and coverage. Output valid JSON only.

\textbf{User Prompt Template.}\\
Task: analyze gaps between seeds and processed outputs, then summarize actionable findings.

\textbf{Seed data (reference examples)}\\
\texttt{\{seed\_data\_samples\}}

\textbf{Processed data (pipeline outputs)}\\
\texttt{\{processed\_data\_samples\}}

Return JSON:
\begin{lstlisting}[basicstyle=\ttfamily\footnotesize,columns=fullflexible,breaklines=true,breakatwhitespace=false,showstringspaces=false]
{
  "has_differences": true,
  "implicit_requirements": [
    {
      "requirement": "...",
      "seed_evidence": "...",
      "processed_gap": "...",
      "severity": "critical|high|medium|low"
    }
  ],
  "style_and_format": {
    "seed_style": "...",
    "processed_style": "...",
    "mismatches": ["..."],
    "severity": "critical|high|medium|low"
  },
  "error_patterns": [
    {
      "type": "format_violation|missing_field|inconsistency|redundancy",
      "pattern": "...",
      "impact": "...",
      "severity": "critical|high|medium|low"
    }
  ],
  "difficulty_and_coverage": {
    "seed_distribution": "...",
    "processed_distribution": "...",
    "mismatch": "...",
    "severity": "critical|high|medium|low"
  },
  "overall_summary": "..."
}
\end{lstlisting}

\vspace{6pt}
\textbf{(2) Experience Mining Prompt ($f_{\text{exp}}$)}

\textbf{System Prompt.}\\
You turn the quality-check findings into compact, actionable experiences that can be appended back to the structured understanding. Each experience must specify: the observed gap, the concrete constraint to add, and how to enforce it in the next orchestration. Each experience must be 30--40 words.

\textbf{User Prompt Template.}\\
Convert the following quality-check JSON into 3--5 experiences.

\textbf{Quality check results}\\
\texttt{\{quality\_check\_result\}}

Write 3--5 numbered experiences. Each experience must be 30--40 words and follow this structure:
(1) context (few words);
(2) gap described using concrete criteria;
(3) a specific constraint and enforcement action phrased as ``To optimize, ...''.

\end{tcolorbox}

\begin{tcolorbox}[
  title=Judge,
  colback=blue!8,
  colframe=blue!55!black,
  arc=1mm,
  boxrule=0.9pt,
  left=1mm,right=1mm,top=1mm,bottom=1mm,
  fonttitle=\scriptsize,
  breakable
]
\footnotesize

You are a data quality assessment expert specializing in deep quality analysis. Analyze the processed data against seed data to identify implicit quality requirements, style consistency, error patterns, and difficulty distribution.

\textbf{Task Description}
\begin{itemize}[leftmargin=1.2em,itemsep=1pt,topsep=2pt]
  \item \textbf{Processed Data}: samples generated by the pipeline from raw data.
  \item \textbf{Seed Data}: reference examples representing expected quality standards and implicit requirements.
  \item \textbf{Goal}: dataset-level comparison (seed and processed samples are not paired by index).
\end{itemize}

\textbf{Analysis Focus}
\begin{enumerate}[leftmargin=1.2em,itemsep=1pt,topsep=2pt]
  \item \textbf{Implicit Quality Requirements}: unstated standards demonstrated by seed data.
  \item \textbf{Style Consistency}: tone/structure/length/language patterns and whether processed data matches.
  \item \textbf{Error Pattern Control}: recurring violations or inconsistencies in processed data.
  \item \textbf{Difficulty Distribution}: complexity and coverage patterns and whether processed data matches.
\end{enumerate}

\textbf{Data Comparison}

\textbf{Seed Data (Reference Examples)}\\
\texttt{\{seed\_data\_samples\}}

\textbf{Processed Data (Pipeline Output)}\\
\texttt{\{processed\_data\_samples\}}

\textbf{Output Requirements}
Output \textbf{only} JSON with the following structure:

\begin{lstlisting}[basicstyle=\ttfamily\footnotesize,columns=fullflexible,breaklines=true,breakatwhitespace=false,showstringspaces=false]
{
  "has_differences": true/false,
  "implicit_quality_requirements": {
    "response_quality_gaps": [
      {
        "requirement": "...",
        "seed_demonstration": "...",
        "processed_gap": "...",
        "severity": "critical|high|medium|low",
        "examples": {"seed_example": "...", "processed_example": "..."}
      }
    ],
    "format_rigor_gaps": [
      {"requirement": "...", "seed_demonstration": "...", "processed_gap": "...",
       "severity": "critical|high|medium|low"}
    ],
    "domain_expertise_gaps": [
      {"requirement": "...", "seed_demonstration": "...", "processed_gap": "...",
       "severity": "critical|high|medium|low"}
    ],
    "completeness_standards": [
      {"requirement": "...", "seed_demonstration": "...", "processed_gap": "...",
       "severity": "critical|high|medium|low"}
    ]
  },
  "style_consistency_analysis": {
    "tone_consistency": {
      "seed_tone": "...", "processed_tone": "...",
      "inconsistencies": ["..."], "severity": "critical|high|medium|low"
    },
    "structure_consistency": {
      "seed_structure_patterns": ["..."],
      "processed_structure_patterns": ["..."],
      "inconsistencies": ["..."], "severity": "critical|high|medium|low"
    },
    "length_distribution": {
      "seed_distribution": "...", "processed_distribution": "...",
      "mismatch": "...", "severity": "critical|high|medium|low"
    },
    "language_patterns": {
      "seed_patterns": ["..."], "processed_patterns": ["..."],
      "inconsistencies": ["..."], "severity": "critical|high|medium|low"
    }
  },
  "error_pattern_analysis": {
    "common_errors": [
      {
        "error_type": "...",
        "description": "...",
        "frequency": "...",
        "trigger_conditions": "...",
        "severity": "critical|high|medium|low",
        "examples": ["..."]
      }
    ],
    "error_distribution": {
      "most_common": "...", "most_critical": "...", "error_rate": "..."
    }
  },
  "difficulty_distribution_analysis": {
    "task_complexity": {
      "seed_distribution": "...", "processed_distribution": "...",
      "mismatch": "...", "severity": "critical|high|medium|low"
    },
    "response_complexity": {
      "seed_distribution": "...", "processed_distribution": "...",
      "mismatch": "...", "severity": "critical|high|medium|low"
    },
    "domain_coverage": {
      "seed_coverage": "...", "processed_coverage": "...",
      "gaps": "...", "severity": "critical|high|medium|low"
    }
  },
  "overall_assessment": "...",
  "critical_insights": ["...", "...", "..."]
}
\end{lstlisting}

If \texttt{has\_differences} is false, it means processed data is basically consistent with seed data and no adjustment is needed.
If \texttt{has\_differences} is true, list the major difference patterns and actionable improvement suggestions in the corresponding fields.

\end{tcolorbox}

\begin{tcolorbox}[
  title=$f_{\text{pipe-evolving}}$: Pipeline-Level Self-Evolving (Experience Generation),
  colback=blue!8,
  colframe=blue!55!black,
  arc=1mm,
  boxrule=0.9pt,
  left=1mm,right=1mm,top=1mm,bottom=1mm,
  fonttitle=\scriptsize,
  breakable
]
\footnotesize

Based on the deep quality analysis results, generate highly specific and actionable experiences that can be appended back to structured understanding for the next round.

\textbf{Input}\\
\texttt{\{quality\_check\_result\}}

\textbf{Key Requirements}
\begin{itemize}[leftmargin=1.2em,itemsep=1pt,topsep=2pt]
  \item Generate \textbf{3--5} experiences covering: implicit requirements, style consistency, error control, and optionally difficulty distribution.
  \item Each experience must be \textbf{exactly 30--40 words}.
  \item Each experience must be \textbf{concrete and actionable}: cite observable criteria (numbers, thresholds, patterns), state the gap, and prescribe an enforceable constraint.
  \item Use the phrasing \textbf{``To optimize, ...''} to introduce the action.
  \item Output \textbf{plain text only} (no JSON).
\end{itemize}

\textbf{Per-Experience Structure (30--40 words total)}
\begin{enumerate}[leftmargin=1.2em,itemsep=1pt,topsep=2pt]
  \item \textbf{Background} (2--5 words): context.
  \item \textbf{Gap} (6--10 words): contrast seed vs processed with measurable criteria.
  \item \textbf{Action} (15--20 words): start with ``To optimize,'' then specify the constraint and where/how to enforce it (e.g., in understanding constraints or orchestration checks).
\end{enumerate}

\textbf{Output Format}
\begin{lstlisting}[basicstyle=\ttfamily\footnotesize,columns=fullflexible,breaklines=true,breakatwhitespace=false,showstringspaces=false]
When processing data, please note the following:

[1]. ...
[2]. ...
[3]. ...
\end{lstlisting}

\end{tcolorbox}

\end{document}